\documentclass[aps,prx,groupedaddress,showpacs,twocolumn]{revtex4-2}
\usepackage[utf8]{inputenc}
\usepackage{hyperref}
\usepackage{graphicx}  
\usepackage{dcolumn}   
\usepackage{bm}        
\usepackage{amssymb,amsmath}   
\usepackage{mathtools}
\usepackage{bibunits}
\usepackage{color}
\usepackage{uri}

\DeclareMathOperator\erf{erf}
\DeclareMathOperator\erfc{erfc}
\DeclareMathOperator\Pe{Pe}

\makeatletter
\newcommand{\hathat}[1]{%
\begingroup%
  \let\macc@kerna\z@%
  \let\macc@kernb\z@%
  \let\macc@nucleus\@empty%
  \hat{\raisebox{.3ex}{\vphantom{\ensuremath{#1}}}\smash{\hat{#1}}}%
\endgroup%
}
\makeatother

\begin{document}

\title{Genome replication in asynchronously growing microbial populations}
\author{Florian Pflug$^1$, Deepak Bhat$^{1,2}$, Simone Pigolotti$^1$}
\email[]{simone.pigolotti@oist.jp}

\affiliation{$^1$Biological Complexity Unit,\\ Okinawa Institute of Science and Technology Graduate University,\\ Onna, Okinawa 904-0495, Japan}

\affiliation{$^2$Department of Physics, School of Advanced Sciences, Vellore Institute of Technology, Vellore, Tamil Nadu, India}

\begin{abstract} 
Biological cells replicate their genomes in a well-planned manner. The DNA replication program of an organism determines the timing at which different genomic regions are replicated, with fundamental consequences for cell homeostasis and genome stability. Qualitatively, in a growing cell culture, one expects that genomic regions that are replicated early should be more abundant than regions that are replicated late. This abundance pattern can be experimentally measured using deep sequencing.  However, a general quantitative theory to explain these data is still lacking. In this paper, we predict the abundance of DNA fragments in asynchronously growing cultures from any given stochastic model of the DNA replication program.  As key examples, we present stochastic models of the DNA replication programs in {\em Escherichia coli} and in budding yeast. In both cases, our approach leads to analytical predictions that are in excellent agreement with experimental data and permit to infer key information about the replication program. In particular, our method is able to infer the locations of known replication origins in budding yeast with high accuracy. These examples demonstrate that our method can provide insight into a broad range of organisms, from bacteria to eukaryotes.
 \end{abstract} 
\maketitle     

\section{Introduction}

The genome of an organism contains precious information about its functioning. Genomes need to be reliably and quickly replicated for cells to pass biological information to the next generation. Replication of a genome proceeds according to a certain plan, termed the ``replication program'' \cite{bechhoefer2012replication,baker2012inferring,gispan2017model,hulke2020genomic}. For example, most bacteria have a circular genome, where two replisomes initiate replication by binding at the same origin site \cite{xu2018bacterial}. They replicate the genome in opposite directions. Replication is completed when they meet, after each of them has copied approximately half of the genome.  The replication program is carefully orchestrated, but not completely deterministic. Stochasticity is particularly relevant in eukaryotes and archaea, where many origin sites are present in each chromosome \cite{gilbert2001making,costa2022initiation}. Replication can initiate at these origin sites at different times. These times are characterized by some degree of randomness \cite{yang2010modeling,rhind2022dna} and, depending on the circumstances, some origin sites may not be activated at all \cite{bechhoefer2012replication}.

Replication programs must be coordinated with the cell cycle in some way. For example, in bacteria, the moment in the cell cycle in which replication is initiated is carefully timed \cite{cooper_chromosome_1968,fu2023bacterial,berger2023synchronous}. In eukaryotes, replication takes place at a well-defined stage of the cell cycle (the S phase), during which different genomic regions are replicated at different times in a controlled manner \cite{yang2010modeling}. The interplay between the replication program and the cell cycle is crucial when trying to infer the replication program from experimental observations. Many experiments study samples in which all cells are approximately at the same stage of the cell cycle. This can be achieved by either arresting the cell cycle at a certain stage or by cell sorting \cite{hulke2020genomic}. The fraction of these synchronized cells that have copied each genome location can be then measured using deep sequencing. This approach has been extensively used to study the eukaryotic replication program \cite{raghuraman2001replication,daigaku2015global,hulke2020genomic}. 

An alternative method is to measure the abundance of DNA fragments in asynchronous, exponentially growing populations. This approach, traditionally called marker frequency analysis \cite{sueoka1965chromosome}, has been extensively applied to bacterial replication, for example to study {\em Escherichia coli} mutant lacking genes that assist DNA replication \cite{wendel2014completion,wendel2018sbcc,midgley2018chromosomal} and artificially engineered {\em E.coli} strains with multiple replication origins \cite{dimude2018origins}. The asynchronous approach is experimentally much simpler and avoids potential artifacts caused by the cell cycle arrest or cell sorting \cite{hulke2020genomic}. Progress in DNA sequencing have made these experiments high-throughput and relatively inexpensive. However, the DNA sampled in these experiments originates from a mixture of cells at different stages in their life cycle, rendering the theoretical interpretation of such data problematic \cite{rhind2010reconciling}. 

Theoretical approaches have attempted to describe measurements in asynchronous populations. However, these approaches either neglect stochasticity, or are limited to specific model systems. For example, we recently proposed a stochastic model describing DNA replication in growing {\em E.coli} populations \cite{bhat2022speed}. A broader range of bacterial systems have been studied assuming a deterministic replication program \cite{huang2022high,huang2022characterizing}. Finally, a model of the replication program in budding yeast adopted the working hypothesis that cells in an asynchronous population are at random, uniformly distributed stages in their cell cycle \cite{gispan2017model}.

In this paper, we develop a general theory to infer the replication program from sequencing of an asynchronously growing population of cells. Our theory builds upon classic results on age-structured populations \cite{powell1956growth,yoshikawa1963sequential,sueoka1965chromosome,chandler1975effect,jafarpour2018bridging}, that we extend to populations of stochastically replicating genomes. Our approach requires minimal assumptions on the replication dynamics. In particular, it allows for a stochastic replication program and equally applies to bacteria and eukaryotes. We apply our method to experimental data from {\em E. coli} and budding yeast. In the case of {\em E. coli}, we propose and exactly solve a model of DNA replication that incorporates stochastic stalling. The fit of the solution to experimental data sheds light on recently observed oscillations of bacterial replisome speed. In the case of yeast, our approach permits to reliably infer the location of replication origins. 

The paper is organized as follows. In Section~\ref{sec:theory}, we present our general theory. In Section~\ref{sec:bacteria}, we apply our framework to DNA replication in {\em E. coli}. In Section~\ref{sec:eukaryotes}, we consider eukaryotic DNA replication, with budding yeast as an example. In Section~\ref{sec:extension}, we present possible extensions of our theory. Section~\ref{sec:discussion} is devoted to conclusions and future perspectives.

\section{DNA replication in asynchronously growing populations}\label{sec:theory}

\subsection{General theory}

\begin{figure*}[htbp]
    \hbox to \textwidth{
    \vtop{\hbox to 0in{\ \textbf{(a)}}\hbox{\includegraphics{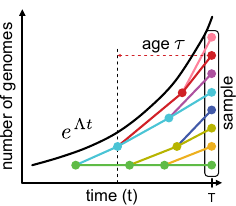}}}%
    \hfill
    \vtop{\hbox to 0in{\ \textbf{(b)}}\hbox{\includegraphics{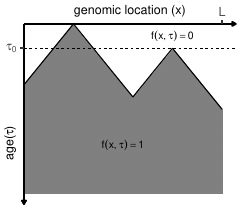}}}%
    \hfill
    \vtop{\hbox to 0in{\ \textbf{(c)}}\hbox{\includegraphics{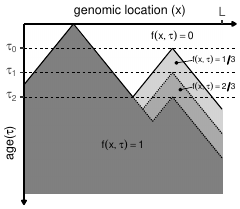}}}%
    \hfill
    \vtop{\hbox to 0in{\ \textbf{(d)}}\hbox{\includegraphics{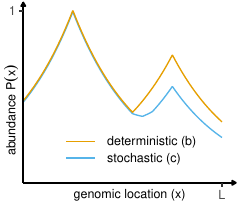}}}%
    }
    \caption{DNA abundance in an asynchronous, exponentially growing microbial population. \textbf{(a)}. Total number of genomes (black line) and genealogy of individual genomes (colored tree). Nodes in the tree represent replication initiations. Such events leave the template unchanged and create a new genome (differently colored descendant) with initial age $\tau=0$. \textbf{(b)}. Example of a deterministic replication program $f(x,\tau)$ on a linear genome in which replication is initiated at two origins, one firing at age $\tau=0$ and one at age $\tau=\tau_0$, and proceeds deterministically with constant speed. \textbf{(c)}. A stochastic version of the replication program in (b), in which replication at the second origin is initiated randomly at either $\tau_0$, $\tau_1$ or $\tau_2$. \textbf{(d)}. DNA abundance distribution arising from replication programs (b) and (c) predicted by Eq.~\eqref{eq:abun1}.}
    \label{fig:overview}
\end{figure*}

We consider a large, growing population of cells. We call $N_c(t)$ the number of cells that are present in the population at time $t$. Each cell may contain multiple genomes, some complete and other undergoing synthesis (incomplete). We denote by $N_g(t)$ the the total number of complete and incomplete genomes in the population.

Our theory is based on the following assumptions. The number of cells grows exponentially: $N_c(t)\propto \exp(\Lambda t)$, where  $\Lambda$ is the exponential growth rate. The population grows in a steady, asynchronous manner. This assumption implies that the average number of genomes per cell must remain constant, and therefore $N_g(t)\propto \exp(\Lambda t)$. Genomes in the population are immortal, i.e. we neglect the rate at which they might be degraded. All genomes in the population are statistically identical, in the sense that they are all characterized by the same stochastic replication program. These assumptions are realistic in common experimental situations. Moreover, as we shall discuss in Section~\ref{sec:extension}, some of them can be readily relaxed, if necessary.

We now assign an age to each genome in the population. To this aim, we conventionally set the birth time of a daughter genome at the start of the replication process that generates it from a parent genome, see Fig.~\ref{fig:overview}a. We note that the distinction between a ``parent'' and a ``daughter'' genome is somewhat arbitrary, as each of them is made up of a preexisting strand and a newly copied complementary one. Since genomes are immortal, the probability density of new genomes is proportional to $\dot{N}_g(t)$, which is proportional to $\exp(\Lambda t)$ as well. It follows that the distribution of ages $\tau$ of genomes in the population must be proportional to $\dot{N}_g(t-\tau)$. From this fact, we conclude that the distribution of genome ages in the population is
\begin{equation}\label{eq:genomeages}
P(\tau)= \Lambda e^{-\Lambda \tau}.
\end{equation}
We now look into the DNA replication program in more details. In bacteria, replication is carried out by a pair of replisomes, that bind at a specific genome site (the origin) and proceed in opposite directions, each replicating both strands. DNA replication is substantially more complex in eukaryotes, where a large number of replisomes replicate the same chromosome, and initiation site might be activated stochastically. We summarize the effect of all these complex processes into the probability $f(x,\tau)$ that the genome location $x$ has already been copied in a genome of age $\tau$. By definition, $f(x,\tau)$ is a non-decreasing function of $\tau$. Our assumption that genomes are statistically identical means that all genomes are characterized by the same $f(x,\tau)$. Examples of deterministic and stochastic replication programs are represented in Fig.~\ref{fig:overview}b and ~\ref{fig:overview}c, respectively.

We now define the probability $\mathcal{P}(x)$ that a randomly chosen genome in the population contains the genome location $x$. We remark that $\mathcal{P}(x)$ is not necessarily normalized to one when integrated over the entire genome. Its normalized counterpart represents the probability density that a randomly chosen genome fragment in the population originates from genomic location $x$. For this reason, we call $\mathcal{P}(x)$ the DNA abundance distribution. The DNA abundance distribution can be experimentally measured using deep sequencing. 

By using Eq.~\eqref{eq:genomeages} and integrating over all genome ages, we find that $\mathcal{P}(x)$ is related with $f(x,\tau)$ by
\begin{equation}\label{eq:abun1}
\mathcal{P}(x)=\int_0^\infty\, d\tau\, \Lambda e^{-\Lambda \tau} f(x,\tau) ,
\end{equation}
see Fig.~\ref{fig:overview}d.
To find a more transparent expression, we introduce the probability density $\psi(x,\tau)=\partial_\tau f(x,\tau)$. Substituting this definition into Eq.~\eqref{eq:abun1} and integrating by parts we obtain
\begin{equation}\label{eq:abun2}
\mathcal{P}(x)=\left\langle e^{-\Lambda \tau} \right \rangle_x \, ,
\end{equation}
where $\langle \dots\rangle_x=\int_0^\infty d\tau \dots \psi(x,\tau)$ is the average over the distribution of replication ages. Equation~\eqref{eq:abun1}, or equivalently Eq.~\eqref{eq:abun2}, is the basis of our approach.

In our derivation, we intentionally avoided modeling the specific dynamics of the cell cycle, how it is regulated, and how it is coordinated with the DNA replication program. Our theory is rigorously valid independently of these aspects, provided that our initial assumptions hold.

\subsection{Deterministic limit}

In the simple case where replication proceeds deterministically and the replisome speed is a function of its position on the DNA, one has
\begin{equation}\label{eq:determ_repl}
\psi(x,\tau)=\delta\left(\tau-\tau_0- \int_{x_0}^x \frac{dx'}{v(x')}\right)
\end{equation}
where $v(x)$ is the replisome speed at position $x$, $x_0$ is the coordinate of the replication origin for the replisome that copied position $x$, and $\tau_0$ is the firing age for that origin. The speed $v$ might take positive or negative values depending on whether the replisome proceeds in the positive genome direction. It follows from Eqs.~\eqref{eq:abun2} and ~\eqref{eq:determ_repl} that
\begin{equation}\label{eq:determ_repl_P}
\mathcal{P}(x)=\exp\left(-\Lambda \tau_0 -\Lambda\int_{x_0}^x \frac{dx'}{v(x')}\right).
\end{equation}
Solving for the speed, we obtain
\begin{equation}\label{eq:v_recovery}
    v(x) = -\Lambda \left[ \frac{d}{dx}\ln \mathcal{P}(x)\right]^{-1} ,
\end{equation}
i.e., the replication speed is inversely proportional to the logarithmic slope of $ \mathcal{P}(x)$ \cite{huang2022characterizing,huang2022high}. An advantage of the deterministic assumption is that it leads to a one-to-one correspondence between DNA abundance and local replisome speed thanks to Eq.~\eqref{eq:v_recovery}. However, in many realistic cases, neglecting stochasticity might lead to inaccurate predictions. 

Unfortunately, in the general stochastic case, different replication programs might give rise to the same DNA abundance distribution. This implies that one can not directly invert Eq.~\eqref{eq:abun2} to express $f(x,\tau)$ in terms of $\mathcal{P}(x)$. In these situations, one has to complement the information contained in $\mathcal{P}(x)$ with modeling assumptions, as exemplified in the following.

\section{Bacterial DNA replication}\label{sec:bacteria}

\begin{figure*}
    \hbox to \textwidth{
    \vtop{\hbox to 0in{\ \textbf{(a)}}\hbox{\includegraphics{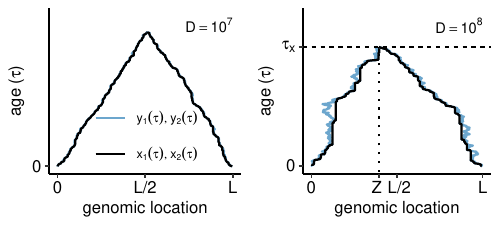}}}%
    \hfill
    \vtop{\hbox to 0in{\ \textbf{(b)}}\hbox{\includegraphics{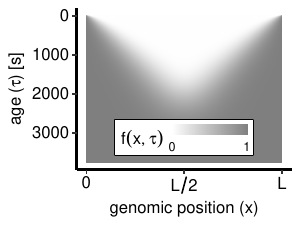}}}%
    \hfill
    \vtop{\hbox to 0in{\ \textbf{(d)}}
    \hbox{\includegraphics{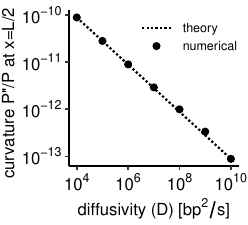}}}%
    }
    \hbox to \textwidth{
    \vtop{\hbox to 0in{\ \textbf{(c)}}
    \hbox{\includegraphics{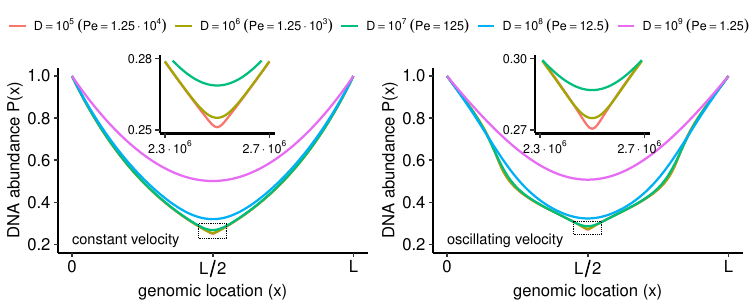}}}%
    \hfill
    \vtop{\vspace{0.3in}\hbox to 0in{\ \textbf{(e)}}\vspace{0.15in}
    \hbox{\includegraphics{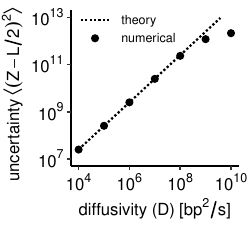}}}%
    }
    \caption{Bacterial DNA replication model. Parameters are: genome length $L=5\cdot 10^6$, growth rate $\Lambda=2\text{h}^{-1}$, baseline speed $v_0=10^3\,\text{bp/s}$. \textbf{(a)}. Trajectories  $x_1$, $x_2$ of the model (black lines) and auxiliary processes $y_1$, $y_2$ (blue lines).
    Backwards movements of $y_1$, $y_2$ correspond to stochastic stalling of $x_1$, $x_2$. Replication concludes at an age $\tau_x$ when $x_1$ and $x_2$ first meet at a random meeting point $Z=x_1(\tau_x)=x_2(\tau_x)$. \textbf{(b)}. Bacterial replication program $f(x,\tau) = 1 - (1-f_1)(1-f_2)$ with $f_1$, $f_2$ from Eq.~\eqref{eq:solfi} for $D=10^8$ $\text{bp$^2$/s}$. \textbf{(c)}. DNA abundance distribution for constant and oscillating replisome speed and different values of $D$. In the constant speed case, $h(\tau) = 1$, whereas in the oscillating speed case $h(\tau) = 1 + \delta\cos(\omega \tau + \phi)$ with $\delta=0.5$, $\omega=2\pi/1800$, and $\phi=0$. \textbf{(d)}. Curvature $\mathcal{P}''(x)/\mathcal{P}(x)$ at the terminus $x=L/2$ predicted by Eq.~\eqref{eq:curvature} vs. numerical simulations. \textbf{(e)}. Uncertainty of the meeting point $Z$ predicted by Eq.~\eqref{eq:uncertainty} vs. numerical simulations.}
    \label{fig:bacterial_model}
\end{figure*}

\subsection{Model}

In this Section, we introduce a stochastic model for the DNA replication of the bacterium {\em E. coli}. The model accounts for variations in the speed of replication, as recently observed \cite{bhat2022speed}. In addition, replisomes in the model can stochastically stall, as observed in single molecule experiments \cite{morin2012active,morin2015mechano}.

Most bacteria, including {\em E. coli}, have a single circular chromosome that is replicated by two replisomes. The two replisomes start from the same origin site, one proceeding clockwise and the other counterclockwise. Replication is completed when the two replisomes meet. We assume that the two replisomes do not backtrack and that they act independently from one another until they meet. A base is therefore replicated by whichever replisome reaches it first. The joint replication program of two replisomes is therefore $f(x,\tau) = 1 - [1 - f_1(x, \tau)][1 - f_2(x, \tau)]$, where $f_1(x,\tau)$ and  $f_2(x,\tau)$ are their individual replication programs, i.e. the probabilities that position $x$ has been replicated at age $\tau$ by the respective replisome. Substituting this expression in Eq.~\eqref{eq:abun1} we obtain
\begin{equation}\label{eq:program_bacteria}
  \mathcal{P}(x) =1-\int_0^\infty d\tau \Lambda e^{-\Lambda \tau}\left(1 - f_1(x, \tau)\right)\left(1 - f_2(x, \tau)\right) .
\end{equation}

We define the genome coordinate $x\in [0,L]$ where $L$ is the genome length. We set the coordinate of the replication origin at $x=0$ (and equivalently $x=L$, since the genome is circular). We call $x_1(\tau)$ and $x_2(\tau)$ the positions of the two replisomes along this coordinate as a function of age. The first replisome starts at $x(0)=0$ and moves in the direction of increasing $x$, whereas the second starts at $x(0)=L$ and moves in the direction of decreasing $x$.

We express the replisome dynamics in terms of two Langevin equations 
\begin{align}\label{eq:langevin}
  \frac{d}{d\tau} y_1 &= v_0 h(\tau) + \sqrt{2Dh(\tau)}\, \xi_1(\tau) \nonumber\\
  \frac{d}{d\tau} y_2 &= -v_0 h(\tau) + \sqrt{2Dh(\tau)}\, \xi_2(\tau) \nonumber\\
  x_1(\tau)&=\max_{\tau'\le \tau} y_1(\tau')\nonumber\\
  x_2(\tau)&=\min_{\tau'\le \tau} y_2(\tau') \ ,
\end{align}
where $\xi_1(\tau)$ and $\xi_2(\tau)$ are Gaussian white noise sources satisfying $\langle\xi_i(\tau)\xi_j(\tau')\rangle=\delta_{ij}\delta(\tau-\tau')$. In Eq.~\eqref{eq:langevin}, we have assumed that the instantaneous average speeds of the two replisomes are opposite in sign. These instantaneous speeds are modulated in time by a function $h(\tau)$, as result of varying conditions during the cell cycle (see \cite{bhat2022speed}). We have assumed that the noise amplitude $\sqrt{2Dh(\tau)}$ is modulated in time  by same the function $h(\tau)$ as the instantaneous average speed. This assumption is also convenient to mathematically solve the model.

\begin{figure*}
\hbox to \textwidth{\vtop{%
\hbox{\vtop{\hbox to 0in{\ \ \textbf{(a)}}\hbox{\includegraphics{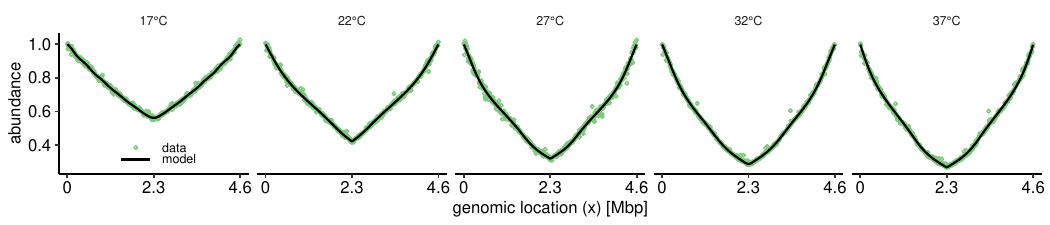}}}}%
\hbox to \textwidth{
\hbox{\vtop{\hbox to 0in{\ \ \textbf{(b)}}\hbox{\includegraphics{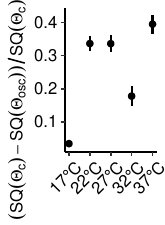}}}}%
\hfill
\hbox{\vtop{\hbox to 0in{\ \ \textbf{(c)}}\hbox{\includegraphics{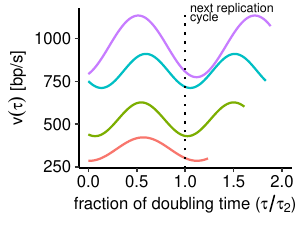}}}}%
\hfill
\hbox{\vtop{\hbox to 0in{\ \ \textbf{(d)}}\hbox{\includegraphics{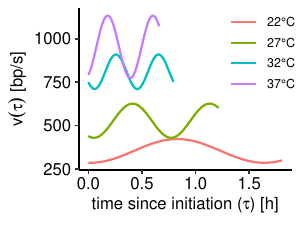}}}}%
\hfill
\hbox{\vtop{\hbox to 0in{\ \ \textbf{(e)}}\hbox{\includegraphics{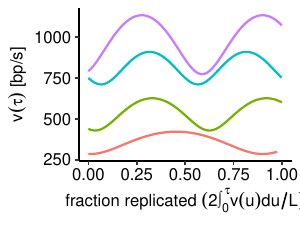}}}}%
}
}}%
\caption{Bacterial model fitted to the data of Ref.~\cite{bhat2022speed}. The replisome speed is modulated by an oscillatory function, see Eq.~\eqref{eq:h_harmonic}. We fitted the parameters $v_0$, $D$, $\delta$, $\omega$, and $\phi$ from the measured DNA abundances. The growth rate $\Lambda$ was independently measured in the experiments (see Appendix \ref{ap:parest}). \textbf{(a)}. Observed DNA abundance and model predictions for E. coli cultures growing at temperatures $T=17^\circ C, 22^\circ C, 27^\circ C, 32^\circ C, 37^\circ C$. \textbf{(b)}. Relative decrease in residuals ($\textrm{SQ}(\Theta) =\sum_{i=1}^N (a_i - \lambda \mathcal{P}(x_i|\Theta))^2 / \sigma_i^2$, see Eq.~\eqref{eq:log-likelihood}) of the model ($\Theta_\textrm{osc}$) vs. the constant speed case ($\Theta_\textrm{c}$). \textbf{(c)} Average instantaneous speed $v(\tau) = v_0 h(\tau)$ as a function of the fraction $\tau / \tau_2$ of the doubling time $\tau_2 = \log(2) / \Lambda$. \textbf{(d)}. Average instantaneous speed $v(\tau) = v_0 h(\tau)$ as the function of the time $\tau$ since replication initiation.\textbf{(e)} Average instantaneous speed $v(\tau) = v_0 h(\tau)$ as a function of the replication progress, i.e. of the fraction $2\int_0^\tau v(u) du / L$ of replicated genome. In (c-e) we omitted $17^\circ$C since the effect of speed fluctuations on $\mathcal{P}(x)$ is negligible at that temperature (see panel b).}
\label{fig:elife_data}
\end{figure*}

The physics of Eq.~\eqref{eq:langevin} can be interpreted as follows. Whenever $y_i(\tau)$ attains a new maximal distance from its origin, then $x_i(\tau)$ is moving forward and it coincides with $y_i(\tau)$. If, instead, $y_i(\tau)$ is making a negative excursion from its past maximal distance, then $x_i(\tau)$ remains frozen at the value of the last maximal distance of $y_i(\tau)$. We interpret these stochastic pausing as replisome stalling events. The amplitude of the noise terms in Eqs.~\eqref{eq:langevin} controls the commonness of long stalling events. Stochastic trajectories generated by the model are qualitatively similar of those observed in single-molecule experiments with DNA polymerases \cite{morin2012active,morin2015mechano}, see Fig.~\ref{fig:bacterial_model}a.

The balance between replisome progress and stalling is governed by the dimensionless Peclet number $\Pe = Lv_0/4D $. For large $\Pe$, the dynamics are nearly deterministic and long stalling events are infrequent. 

A consequence of Eq.~\eqref{eq:langevin} is that the individual replication program $f_i(x,\tau)$ is equal to the first-passage probability of the associated process $y_i$ through $x$. This first-passage probability is expressed by
\begin{align}\label{eq:solfi}
    f_i(x, \tau) &= \int_0^{H(\tau)}
      \sqrt{\frac{\mu_i^3}{2\pi\sigma_i^2 u^3}}
      \exp\left(-\frac{1}{2}\frac{(u-\mu_i)^2}{\sigma_i^2}
                \frac{\mu_i}{u}\right)
    \,du \, ,
 &   \nonumber\\
\end{align}
where $H(\tau)= \int_0^\tau h(u) du$ and
\begin{align}\label{eq:solfi_params}
    \mu_1 &= x/v_0, &\mu_2 &= (L-x)/v_0, \nonumber \\
    \sigma_1^2 &= 2D x / v_0^3, &\sigma_2^2 &= 2D (L-x) / v_0^3\ . \end{align}
Eqs.~\eqref{eq:solfi} and \eqref{eq:solfi_params} are derived in Appendix~\ref{app:bacteriatimedepend}. We compute the DNA abundance $\mathcal{P}(x)$ by substituting Eq.~\eqref{eq:solfi} into Eq.~\eqref{eq:program_bacteria}, see Fig.~\ref{fig:bacterial_model}b. We numerically evaluate the final integral over $\tau$ appearing in Eq.~\eqref{eq:program_bacteria}. 

The main effect of diffusivity is to smoothen the DNA abundance distribution predicted by the model, see Fig.~\ref{fig:bacterial_model}c. This effect is particularly pronounced at the expected meeting point $x=L/2$ of the two replisomes, where the DNA abundance exhibits a cusp for $D=0$ but is smooth for positive $D$, see Figure~\ref{fig:bacterial_model}c. 
In contrast, far from the meeting point diffusivity does not affect the DNA abundance much, see Appendix \ref{ap:curv}. We now quantify the effect of diffusivity by means of Eq.~\eqref{eq:solfi}. In the case of constant speed $v(\tau)=v_0$, the relative curvature of $\mathcal{P}(x)$ at the expected meeting point is approximated by
 \begin{equation}\label{eq:curvature}
    \frac{\mathcal{P}''(L/2)}{\mathcal{P}(L/2)} \approx \frac{2\Lambda}{\sqrt{\pi DL v_0}}\ ,
\end{equation}
see Fig.~\ref{fig:bacterial_model}d and Appendix \ref{ap:curv}. In principle, Eq.~\eqref{eq:curvature} can be used to estimate the diffusion coefficient from the curvature in the terminus region, without having to fit the entire distribution.

From the point of view of trajectories, the smoothing of $\mathcal{P}(x)$ for $D > 0$ occurs because the two replisomes no longer necessarily meet exactly at $x=L/2$. In particular, we find that the uncertainty on the location $Z$ of the meeting point is approximately equal to
\begin{equation}\label{eq:uncertainty}
    \left\langle (Z-L/2)^2 \right\rangle \approx \frac{DL}{2v_0}
    = \frac{L^2}{8\Pe}\ ,
\end{equation}
see Fig.~\ref{fig:bacterial_model}e. Equation~\eqref{eq:uncertainty} is derived in Appendix \ref{ap:meetingpoint}.

\subsection{Comparison with E.coli sequencing data}

\begin{table*}
    \centering
    \begin{tabular}{cccccccc}
        $T$ [$^\circ$C] & $v_0$ [$\text{bp}/\text{s}$] & $D$ [$\text{kbp}^2/\text{s}$] &
        $\delta$ & $\omega$ [\text{rad}/\text{h}] & $\phi$ [rad] &
        $\Pe$ & $\Delta Z$ [kbp]
        \\
        \colrule
        17 & 230 $\pm$ 30 & 1.6 $\pm$ 1.4   & 0.29 $\pm$ 0.1   & 5.8 $\pm$ 4.2  & 1.3 $\pm$ 1.5
         & 100 $\pm$ 100 & 130 $\pm$ 50 \\
        22 & 350 $\pm$ 21 & 0.5 $\pm$ 0.8 & 0.19 $\pm$ 0.07 & 3.9 $\pm$ 0.6
         & 3.1 $\pm$ 0.5
         & 200 $\pm$ 30 & 110 $\pm$ 7 \\
        27 & 530 $\pm$ 20 & 1.0 $\pm$ 1.2  & 0.19 $\pm$ 0.03 & 8.7 $\pm$ 0.4
         & 2.0 $\pm$ 0.2
         & 400 $\pm$ 400 & 80 $\pm$ 40 \\
        32 & 810 $\pm$ 57 & 1.7 $\pm$ 1.8   & 0.12 $\pm$ 0.04  & 16 $\pm$ 1.3   & 1.5 $\pm$ 0.2
         & 500 $\pm$ 400 & 80 $\pm$ 30 \\
        37 & 950 $\pm$ 24 & 1.2 $\pm$ 2.3   & 0.19 $\pm$ 0.03 & 15 $\pm$ 0.4    & 2.9 $\pm$ 0.2
         & 500 $\pm$ 600 & 70 $\pm$ 50 \\
    \end{tabular}
    \caption{Parameter estimates for time-dependent speed $v(t) = v_0\big(1 + \delta\cos(\omega t + \phi)\big)$. For $v_0$, $D$, $\delta$, $\omega$, $\phi$ the reported standard errors represent the variability over replicates. The Peclet number $Pe = L v_0 / 4D$ and meeting point uncertainty $\Delta Z = \sqrt{DL / 2 v_0}$, see Eq.~\eqref{eq:uncertainty}, are computed from the average estimates of $v_0$ and $D$ over replicates where $D > 0$. Their standard error are estimated using error propagation.}
    \label{tab:params_vt}
\end{table*}

We now fit our model of replisome dynamics to experimentally measured DNA abundance from wild type {\em E. coli} grown at different temperatures (from Ref.~\cite{bhat2022speed}). We assume that the speed and diffusion coefficient are modulated in time by a function
\begin{equation}\label{eq:h_harmonic}
h(\tau)= 1 + \delta \cos(\omega \tau + \phi),
\end{equation}
see Fig.~\ref{fig:bacterial_model}. In the fit, we treat $v_0, D, \delta, \omega$, and $\phi$ as free parameters (see Appendix \ref{ap:parest}).

The model fits the experimental DNA abundances very well, see Fig.~\ref{fig:elife_data}a. The estimates of the mean speed $v_0$ are highly consistent among replicates across all temperatures, see Table \ref{tab:params_vt}. At temperatures above $17^\circ$C we find robust evidence of speed fluctuations. The model provides consistent estimates of $\delta$, $\omega$ and $\phi$ in these cases, with an improvement of the quality of fit ranging between 20\% to 40\% percent compared to the constant-speed case (Fig.~\ref{fig:elife_data}b). At $17^\circ$C, the effect of the speed fluctuations on $\mathcal{P}(x)$ is small and, consequently, the uncertainty in the associated parameters $\omega$ and $\phi$ is high. Regardless of temperature, model selection appears to prefer a vanishing value of $D$ for some replicates. The reason is likely that the estimates of $D$ correspond to Peclet numbers  in the range from $\Pe\approx200$ to $\Pe\approx1000$, which lies close to the detection threshold; see Appendix \ref{ap:parest}.

The frequency of speed oscillations appears linked with the population doubling time. In fact, the oscillations for $22^\circ$C to $37^\circ$C align well when time is rescaled according to the duration of a cell cycle, see Fig.~\ref{fig:elife_data}c. The speed consistently attains a minimum after one doubling time $\tau_2 = \log(2) / \Lambda$ when the next replication cycle starts and the number of active forks is thus increased. As comparisons, when plotting the oscillations against absolute time since replication initiation (Fig.~\ref{fig:elife_data}d) or against replication progress (Fig.~\ref{fig:elife_data}e) the alignment is substantially worse. 

The speed oscillations were first observed and quantified using a model in which speed is modulated in space, rather than in time \cite{bhat2022speed}. Our approach permits to analytically solve also a spatially modulated speed model in the small-noise limit. The resulting parameter values well match those of Ref. \cite{bhat2022speed}, and are consistent with our temporally modulated model, see Appendix~\ref{ap:vx} for details. The model with temporal speed oscillations yields a better fit for the majority of samples, consistently with the idea that the speed variations are linked with the doubling time. The improvement in likelihood is, however, small (Fig.~\ref{fig:elifecmp_osc}b in Appendix~\ref{ap:vx}).

\section{Eukaryotic DNA replication}\label{sec:eukaryotes}

\subsection{Model}

\begin{figure*}
\vbox{%
\vtop{\hbox to \textwidth{%
        \hbox{\vtop{\hbox to 0in{\ \ \textbf{(a)}}
                    \hbox{\includegraphics{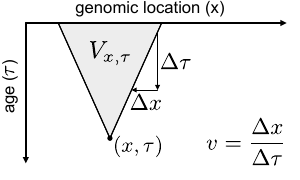}}}}%
        \hfill
        \hbox{\vtop{\hbox to 0in{\ \ \textbf{(b)}}
                    \hbox{\includegraphics{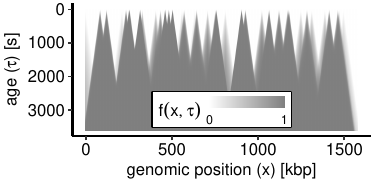}}}}%
        \hfill
        \hbox{\vtop{\hbox to 0in{\ \ \textbf{(c)}}
                    \hbox{\includegraphics{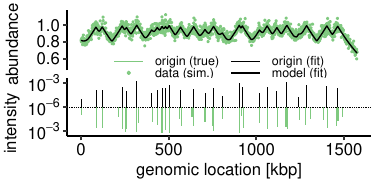}}}}%
}}%
}%
\caption{Eukaryotic model. \textbf{(a)}. Past light-cone $V_{x,\tau}$ of a space-time point $(x,\tau)$. At least one origin must have fired within the light cone for location $x$ to be replicated by time $\tau$. \textbf{(b)}. Eukaryotic replication program from Eq.~\eqref{eq:abundance_eukaryotic} for the annotated origins on \textit{S. cerevisiae} W303 chromosome IV with intensities $I^\star_i / v$ randomly log-uniformly distributed on [$10^{-5}$, $2\cdot 10^-3$]. \textbf{(c)}. Inference of origin locations and intensities from a simulated DNA abundance. The top plot shows the model (black line) fitted to simulated abundances (green line). The bottom plot shows the inferred 26 origins (black bars) and 39 true origins (green bars). Rescaled ntensities $I^\star_i/v$ [$1/\text{bp}$] are plotted in log-scale. Parameters are: $v=27\ \text{bp/s}$, $\Lambda = 9.6\cdot 10^{-5}\ 1/\text{s}$ (doubling time 120 minutes). The true origin locations and intensities are those from panel (b). The resulting true abundances are multiplied by a Gamma-distributed random variable with mean 1.0 and coefficient of variation 0.04 to mimic measurement errors.}
\label{fig:eukaryote}
\end{figure*}

In this Section, we apply our approach to eukaryotic DNA replication. The eukaryotic case is substantially more complex than the bacterial one, because of the presence of multiple, randomly activated replication origins.  Also in this case, when an origin fires, two replisomes start moving from that origin in opposite directions.  In this case, theoretical progress \cite{jun2005nucleation,jun2005nucleation2} has made use of an analogy with the physics of freezing/crystallization kinetics, as described by the so-called Kolmogorov-Johnson-Mehl-Avrami model \cite{kolmogorov1937statistical,william1939reaction,avrami1939kinetics}, see also \cite{sekimoto1986evolution,sekimoto1991evolution}. We briefly summarize this idea and then extend it to asynchronously growing populations.

To fix the ideas, we consider a given location $x$ at an age $\tau$ in a eukaryotic genome. A main problem of the eukaryotic case is that, in principle, a location $x$ could have been replicated by different replisome starting from different origins. The replication program $f(x,\tau)$ can be seen as the probability that the past ``light-cone'' $V_{x,\tau}$ of the space-time point $x,\tau$ contains at least one origin firing event, see Fig.~\ref{fig:eukaryote}a. This elegant argument circumvents the problem of determining from which origin did the replisome that replicated the genome location $x$ start.

Assuming that replisomes progress deterministically with constant speed $v_0$, the past light-cone is the set of space-time points $x',\tau'$ from which $x$ is reachable by a replisome within a time $\tau$, i.e.,
\begin{equation}\label{eq:lightcone}
    V_{x,\tau} = \big\{(x',\tau') \,\big|\, |x - x'|\leq v_0\left(\tau - \tau'\right) \big\},
\end{equation}
see Fig.~\ref{fig:eukaryote}a.  Following \cite{bechhoefer2012replication,baker_inferring_2014}, we now assume that origins fire independently from one another in space and time with position- and age-dependent rates $I(x,\tau)$. This assumption implies that the space-time points at which origins fire form a two-dimensional Poisson point process.

The probability that location $x$ has been replicated by age $\tau$ is then expressed by
\begin{equation}\label{eq:program_cone}
    f(x,\tau) = 1 - \exp\left(-\iint_{V_{x,\tau}} dx'd\tau'\, I(x',\tau')\right) .
\end{equation}
We note that Eq.~\eqref{eq:program_cone} would remain valid if we chose a different replisome dynamics, corresponding to a form of the light-cone $V_{x,\tau}$ other than that expressed by Eq.~\eqref{eq:lightcone}. Using Eq.~\eqref{eq:abun1}, we formally express the DNA abundance distribution as
\begin{equation}\label{eq:abundance_cone}
    \mathcal{P}(x) = 1 - \Lambda \int_{\mathrlap{0}}^{\mathrlap{\infty}}
                     d\tau 
                     \exp\left(-\Lambda \tau-\iint_{\mathrlap{V_{x,\tau}}} dx'd\tau'\, I(x',\tau')\right).
\end{equation}

We now focus on budding yeast, where origins correlate with specific DNA motifs and are therefore thought to be at well-defined locations $x_1,\ldots,x_K$ on the chromosomes \cite{gilbert2001making}. We therefore assume $I(x,\tau) = \sum_{j=1}^{K} I_j(\tau) \delta(x - x_j)$.  We further assume that firing rates are constant in time, so that
\begin{equation}
    I(x,\tau) = \sum_{j=1}^K I^\star_j \delta(x - x_j).
    \label{eq:yeast_Ixt}
\end{equation}
In budding yeast, the origin firing rate was observed to be time-dependent \cite{yang2010modeling}. This means that this second assumption is not fully accurate and should be considered as a simplifying approximation.

To express the DNA abundance $\mathcal{P}(x)$ for firing rates as in Eq.~\ref{eq:yeast_Ixt}, we consider the travel time $\tau_k$ from origin $x_k$ to a given position $x$,
\begin{equation}
  \tau_k = \frac{\left|x - x_k\right|}{v_0}.
\end{equation}
At fixed $x$, we reorder the origins so that $0 < \tau_1 < \cdots < \tau_K$. We then define the effective replication times
\begin{equation}
  \mathcal{T}_k = \tau_k + \sum_{j=1}^k \left(\tau_k - \tau_j\right) I^\star_j / \Lambda
  \label{eq:eukaryotes_Ti}
\end{equation}
which collectively take into account the effect of all origins. We also introduce the effective cumulative weights
\begin{equation}
  W_k = 1 + \sum_{j=1}^k I^\star_j / \Lambda ,
  \label{eq:eukaryotes_Wi}
\end{equation}
with the convention that $W_0=1$.
Using these definitions, we express the DNA abundance by
\begin{equation}
  \mathcal{P}(x) =
  \sum_{k=1}^K \left(\frac{1}{W_{k-1}} - \frac{1}{W_k}\right)e^{-\Lambda \mathcal{T}_k} .
  \label{eq:abundance_eukaryotic}
\end{equation}Equation~\eqref{eq:abundance_eukaryotic} is derived in Appendix \ref{ap:eukaryotes}.

We implemented a simulated annealing algorithm to infer the origin number, location, and intensities based on DNA abundance data via Eq.~\eqref{eq:abundance_eukaryotic}. Our inference procedure treats $\Lambda / v_0$, the number of origins $K$, the origin positions $x_1,\ldots,x_K$, and the re-scaled firing rates $I^\star_1/v_0,\ldots,I^\star_K/v_0$ as free parameters. We call the compound parameter $I^\star_j/v_0$ the intensity of origin $j$. Our algorithm uses the Akaike information criterion (AIC) as the cost function to avoid over-fitting (details in Appendix \ref{ap:simann}).

We first tested this inference procedure on artificial data, in which the genome length (1.6Mbp) and origins distribution are comparable to those of the longest chromosome of budding yeast (chromosome IV). Our inference algorithm detected the location and intensity of 26 out of 39 true origins with high accuracy (Fig.~\ref{fig:eukaryote}c; median distance to true origin 1.5kb, median relative error of intensity 40\% if non-resolved clusters of true origins are merged; true intensities range over two orders of magnitude). The 13 non-recovered origins either have low intensity or were merged with another origin in close proximity.

\subsection{Application to experimental data}

\begin{figure*}
\vbox{%
    \vtop{\hbox{\vtop{\hbox to 0in{\ \ \textbf{(a)}}
                      \hbox{\includegraphics{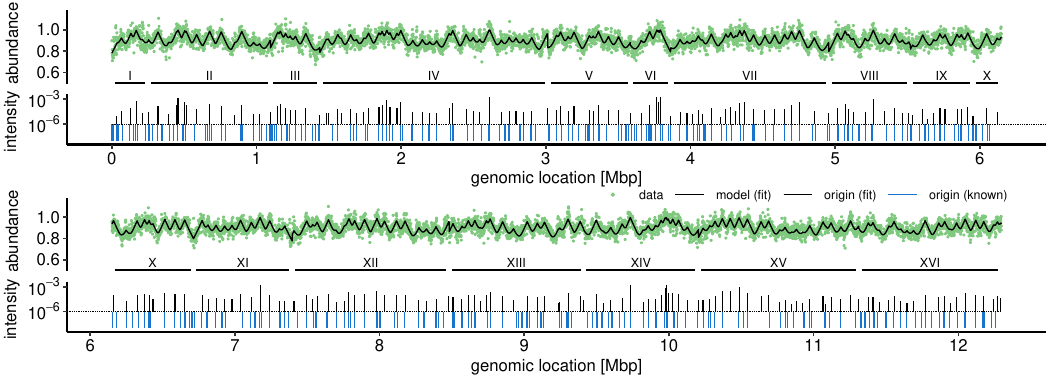}}}}}%
    \vtop{\hbox to \textwidth{%
        \hbox{\vtop{\hbox to 0in{\ \ \textbf{(b)}}
                    \hbox{\includegraphics{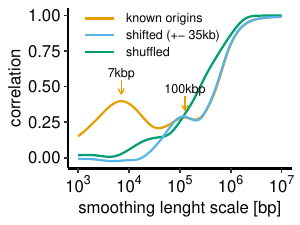}}}}%
        \hfill%
        \hbox{\vtop{\hbox to 0in{\ \ \textbf{(c)}}
                    \hbox{\includegraphics{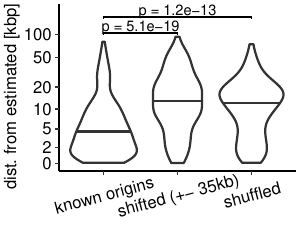}}}}%
        \hfill%
        \hbox{\vtop{\hbox to 0in{\ \ \textbf{(d)}}
                    \hbox{\includegraphics{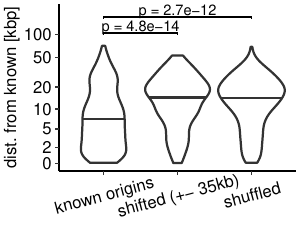}}}}%
    }}%
}%
\caption{Eukaryotic model fitted to \textit{S. cerevisiae} data from \cite{muller_2013}. Fitted parameters are: $\Lambda/v$, origin positions $x_1,\ldots,x_n$ and intensities $I^\star_1/v,\ldots,I^\star_n/v$. The fitted number of origins is $K=234$. Known origins used for validation  were lifted from the annotated genome for strain 288C (RefSeq assembly \textit{R64-3-1}, accession \textit{GCF\_000146045.2}) using \textit{liftoff} \cite{shumate_liftoff_2021}. We excluded the mitochondrial genome. \textbf{(a)}. Observed (green circles) and predicted (black line) DNA abundances (top) and inferred origin positions and intensities (bottom). Known origin positions (blue) shown for comparison. \textbf{(b)}. Correlation of estimated ($K=234$) and known ($K=354$) origin densities at different smoothing length scales (yellow). For comparison, we also show the correlation to the density of shuffled (green) and randomly shifted (blue) known origins. \textbf{(c)}. Distribution of distances between estimated and closest known origins. \textbf{(d)}. Distribution of distances between known and closest estimated origins. In (c) and (d), the y-axis is non-linearly scaled to enlarge the region of interest. Horizontal lines indicate the medians. Reported p-values are computed using Mann–Whitney U tests.}
\label{fig:yeast}
\end{figure*}

We now apply our model and inference procedure to experimentally measured DNA abundance in an asynchronous populations of budding yeast (\textit{S. cerevisiae} W303) \cite{muller_2013}. Our algorithm infers $K=234$ origins across the 16 yeast chromosomes. The predicted DNA abundance well matches the experimental one, see Fig.~\ref{fig:yeast}a.

Our method well predicts origins of replication with a resolution on the order of a few kilobases, and without using additional experimental information other than the DNA abundance distribution. The inferred origins locations correlate with known origins of \textit{S. cerevisiae} when binned on a length scale of about 7kb, see Fig.~\ref{fig:yeast}b. A second correlation peak at 100kb suggests a large-scale pattern in the distribution of origins. As expected, the peak at 7kb vanishes when known origin locations are randomly shifted by $\pm 35\text{kbp}$, or shuffled by randomly reordering chromosomes. The results of the correlation analysis are corroborated by matching each inferred origin to the closed known origin (figure \ref{fig:yeast}c; median distance 3.9kb) respectively each known origin to the closest inferred origin (figure \ref{fig:yeast}d; median distance 6.6kb). In both cases, the average distances are significantly increased if the known origins are shifted or shuffled.

\section{Extensions of the theory}\label{sec:extension}

\subsection{Non-exponential growth}

Our approach can be generalized to populations that do not grow exponentially. We here assume that the number of genomes grows with time according to an arbitrary function $N_g(t)$. In this case, following the same logic of Section~\ref{sec:theory}, we find that Eq.~\eqref{eq:abun1} becomes
\begin{equation}\label{eq:abun_nonexp}
    \mathcal{P}(x,t) = \int_0^\infty d\tau P(\tau;t) f(x,\tau)
\end{equation}
where $P(\tau;t)\propto N'_g(t-\tau)$. The proportionality constant can be determined by imposing normalization of $P(\tau)$. After integrating Eq.~\ref{eq:abun_nonexp} by parts, we obtain
\begin{equation}\label{eq:abun_nonexp2}
    \mathcal{P}(x,t) = \left \langle N_g(t-\tau) \right\rangle .
\end{equation}
This generalization can be applied, for example, to bacterial or eukaryotic populations that grow in a partially synchronous manner. In such cases, $N_g(t)$ might exhibit temporal peaks. 

We note that, in Eqs.~\eqref{eq:abun_nonexp} and ~\eqref{eq:abun_nonexp2} we implicitly made the assumption that the replication program $f(x,\tau)$ does not depend on $t$. The validity of this assumption should be verified for populations that are out of steady growth. Moreover, a practical limiting factor can be the accuracy at which $N_g(t)$ can be estimated.

\subsection{Non-immortal genomes}

For the purpose of computing the DNA abundance distribution, our assumption of immortal genomes is justified as long as typical genome lifetimes $\tau_\text{max}$ are much longer than the doubling time $\tau_2 = \log(2) / \Lambda$. This is usually a realistic assumption, but in some cases one may want to generalize our approach to include genome degradation. In such case, the exponential distribution in Eq.~\eqref{eq:abun1} should be replaced with the age distribution $P(\tau)$ of surviving genomes. In particular, if genomes survive until age $\tau$ with probability $s(\tau)$, the age distribution is expressed by
\begin{equation}
    P(\tau) \propto  \Lambda e^ {-\Lambda \tau} s(\tau)
\end{equation}
up to a normalization constant.

\subsection{Multiple genome types}

The theory we developed so far applies to statistically identical genomes. This assumption is sometimes referred to as ``identity at birth'' in the literature on renewal processes \cite{boldin2023population}: all genomes are born identical, and the differences that they might develop as they synthesize can be seen as independent outcomes of the same stochastic process. 

There might be cases in which this assumption does not hold. For example, certain knockout mutations of {\em E. coli} lead to defective genome types that are inheritable \cite{sinha2018broken}. To address these scenarios, we extend our theory to cases in which different genome types coexist in the population, each characterized by a different replication program.

We call $N_g^{(i)}(t)$ the number of genomes of type $i$ at time $t$ in the population, with $i=1\dots k$. The total number of genomes is $N_g(t)=\sum_i N_g^{(i)}(t)$. Each genome of type $i$ can be replicated into a genome of type $i'$ according a certain stochastic process. We assume that this stochastic process satisfies the Perron-Frobenius theorem, so that at long time the number of genomes are given by
\begin{equation}
N_g^{(i)}(t) \propto r_i\, e^{\Lambda t} .
\end{equation}
Here $\Lambda$ is the leading eigenvalue of the dynamics, corresponding to the exponential growth rate of the population. The vector $(r_1\dots r_k)$ is the leading eigenvector, having non-negative real entries according to the Perron-Frobenius theorem, and normalized such that $\sum_i r_i =1$. The resulting replication program can then be obtained as an average over the genome types:
\begin{equation}
  f(x,\tau) = \sum_{i=1}^k r_i f^{(i)}(x,\tau).
\end{equation}
This average replication program can be substituted into Eq.~\eqref{eq:abun1} to determine the DNA abundance distribution.

\subsection{Numerical simulations and analogy with stochastic resetting}

The models we considered in Sections \ref{sec:bacteria} and \ref{sec:eukaryotes} are analytically solvable. However, this might not be the case for more complex models. In this Section, we briefly present an interpretation of the DNA abundance distribution $\mathcal{P}(x)$ as the steady-state of a dynamical process implementing stochastic resetting \cite{evans2011diffusion,evans2020stochastic}. This analogy can be used for efficient numerical simulations.

The models introduced in Sections \ref{sec:bacteria} and \ref{sec:eukaryotes} are described by linear equations. Formally, we assume that we can always write the stochastic process associated with a certain DNA replication program in terms of linear evolution operator $\hat{\mathcal{L}}$, so that
\begin{equation}\label{eq:program_dynamics}
\partial_\tau f(x,\tau) = \hat{\mathcal{L}} f(x,\tau). 
\end{equation}
We now introduce a modified version of this equation:
\begin{equation}\label{eq:program_dynamics_resetting}
\partial_\tau f(x,\tau) = \hat{\mathcal{L}} f(x,\tau) - \Lambda f(x,\tau). 
\end{equation}
Equation~\ref{eq:program_dynamics_resetting} can be interpreted as describing a replication program equivalent to that of Eq.~\eqref{eq:program_dynamics}, but that is stochastically reset to its beginning at rate $\Lambda$. Stochastic simulations of this process can be easily implemented numerically. 

We note that $\mathcal{P}(x)$ is a steady solution of Eq.~\eqref{eq:program_dynamics_resetting}, i.e., it satisfies
\begin{equation}\label{eq:resetting}
\left(\hat{\mathcal{L}} - \Lambda\right) \mathcal{P}=0.
\end{equation}
This means that we can determine $\mathcal{P}(x)$, up to an appropriate normalization constant, by sampling 
the stochastic dynamics described Eq.~\eqref{eq:program_dynamics_resetting} at steady state. Such stochastic simulations can be efficiently implemented numerically \cite{bhat2022speed}.

\section{Conclusions}\label{sec:discussion}

In this paper, we introduced a general theory that connects the DNA replication program with the abundance of DNA fragments that one should expect in an asynchronously growing population of cells. Our theory builds on previous approaches \cite{yoshikawa1963sequential,sueoka1965chromosome,gispan2017model,bhat2022speed,huang2022high} and has the advantage of being based on a minimal set of realistic assumptions and allowing for stochastic replication programs. As we have demonstrated, these key properties make our theory applicable to a broad range of organisms, from bacteria to eukaryotes. 

In the case of bacterial DNA replication, we proposed a model in which the replisome speed is modulated in time and replisomes can stochastically stall. We solved the model exactly and fitted its prediction against sequencing data of {\em E.coli} growing at different temperatures \cite{bhat2022speed}. The fits show that the period of speed oscillations matches the population doubling time, or equivalently the time interval between consecutive fork firing. Our model with time-periodic speed variations fits the data slightly better than the one with space-periodic variations as postulated in \cite{bhat2022speed}. Taken together, these observations support that the causes of oscillations are linked with the cell cycle, or alternatively with the fork firing rate. A possible candidate would be competition among multiple forks on the same genome \cite{bhat2022speed}. At variance with Ref.~\cite{bhat2022speed}, the approach introduced in this paper leads to an analytical expression for the DNA abundance distribution, which considerably simplifies the inference procedure and provides additional physical insight.

The fits of the {\em E.coli} data reveal that the Peclet number characterizing the replisome dynamics is rather large. On the one hand, this observation confirms that simpler approaches that neglect stochasticity \cite{huang2022characterizing,huang2022high} provide reliable results, at least in the case of wild type {\em E.coli}. On the other hand, the diffusion constant, albeit small, provides important information about the uncertainty of the replisome meeting point. In {\em E.coli}, the Tus-Ter system is know to set bounds on the region in which replisomes can meet \cite{elshenawy2015replisome}, thereby likely affecting this accuracy. It will be interesting for future studies to apply our approach to mutant strains, to see whether they are characterized by a different degree of uncertainty.

Stochasticity is definitely crucial in the eukaryotic case. We have used our approach to estimate the origins location and intensities for budding yeast from the DNA abundance distribution measured in \cite{nieduszynski_2006}. Our approach is based on seminal work by Bechhoefer and coworkers \cite{bechhoefer2012replication,baker_inferring_2014}, that we extended to asynchronously growing populations.  A previous study \cite{gispan2017model} also attempted at extending the approach from \cite{bechhoefer2012replication,baker_inferring_2014} to asynchronously growing budding yeast. Our results differ from those of Ref.~\cite{gispan2017model} in two different aspects. First, Ref.~\cite{gispan2017model} assumed as a working hypothesis a uniform distribution for the distribution $P(\tau)$. In contrast, our central result in Section \ref{sec:theory} shows that the distribution $P(\tau)$ should be exponential under very general conditions. Second, in fitting the model to the data, Ref.~\cite{gispan2017model} used experimental knowledge of the origin coordinates. Instead, our method was able to directly infer these coordinates, without requiring any additional experimental information other than the DNA abundance distribution.

In our eukaryotic model, we assumed for simplicity that replisome speed is constant; that origins are placed at well defined sites; and that they fire at an origin-dependent rate that is constant in time. The last assumption, in particular, is a drastic approximation, since origin firing rates in yeast are known to be markedly time-dependent \cite{yang2010modeling}. Relaxing these assumptions constitutes an important challenge for future research and will permit to recover origin timing behaviour, beside locations, and thus provide a more complete picture of the replication program. 

In any case, despite these simplifying assumptions, our algorithm successfully recovers the locations of the majority of known origins in budding yeast, with an accuracy on the order of kilobases. The accuracy can likely be further increased by exploiting advances in sequencing technology, in particular increased sequencing depth and read lengths, and by further improving the optimization algorithm. Our results demonstrate that the combination of deep sequencing of asynchronous populations and our inference approach provides a cost-effective way of discovering the replication origins of any single-cell eukaryotic species which can be cultured and sequenced.

\begin{acknowledgments}
We thank S. Hauf, C. Plessy, and Y. Yokobayashi for fruitful discussions. We thank A. Alsina, J. Bechoefer, N. Rhind, P. Sartori,  and A. Sassi for feedback on a preliminary manuscript. This work was supported by JSPS KAKENHI Grant No. 23H01146 (to SP). DB thanks Vellore Institute of Technology, Vellore for providing VIT SEED Grant-RGEMS Fund (SG20220060) for carrying out this research work.
\end{acknowledgments}

\appendix

\section{Bacterial DNA replication with time-dependent speed}\label{app:bacteriatimedepend}

We consider the Langevin Eqs.~\eqref{eq:langevin}. We write the associated Fokker-Planck equation for the first replisome:
\begin{equation}\label{eq:bacterial_model_fpe}
  \frac{\partial}{\partial \tau} p = h(\tau)\left(-v_0\frac{\partial}{\partial y} p + D\frac{\partial^2}{\partial y^2} p\right).
\end{equation}
Equation~\eqref{eq:bacterial_model_fpe} shows that the function $h(\tau)$ acts as global time rescaling. Therefore, if $\hat p$ is a solution of the Fokker-Planck equation with constant drift and diffusion terms
\begin{equation}\label{eq:drift_diffusion_fpe}
  \frac{\partial}{\partial \tau} \hat p = -v_0\frac{\partial}{\partial y} \hat p + D\frac{\partial^2}{\partial y^2} \hat p
\end{equation}
under certain boundary and initial conditions, then 
\begin{align}
    p(y, \tau) &= \hat p\big(y, H(\tau)\big) 
\end{align}
solves the original Fokker-Planck Eq.~\eqref{eq:bacterial_model_fpe} with the same boundary and initial conditions, where $H(\tau)=\int_0^\tau du\,h(u)$.  We impose, in particular, an absorbing boundary condition at $x$, $\hat p(x, \tau) = 0$ $\forall \tau$ and an initial condition $p(y,\tau=0)=\delta(y)$. The flux $\hat J = v_0 p - D\partial_y p$ through the absorbing boundary is given by an inverse Gaussian distribution:
\begin{equation}
    \hat J(x, \tau) = \frac{x}{\sqrt{4\pi D \tau^3}} e^{-\frac{(x - v_0 \tau)^2}{4D\tau}}.
\end{equation}
Since the first-passage density of Eq.~\eqref{eq:bacterial_model_fpe} equals the flux $J(x,\tau) = h(\tau) \hat J\big(x,H(\tau)\big)$, we have  
\begin{equation}\label{eq:programs_beforeint}
        \frac{\partial}{\partial\tau} f_1(x,\tau) = h(\tau) \hat J\big(x,H(\tau)\big) = \frac{h(\tau)x}{\sqrt{4\pi D H^3(\tau)}} e^{-\frac{(x - v_0H(\tau))^2}{4DH(\tau)}}.
\end{equation}
The same approach can be used for the second replisome after a suitable change of coordinates and initial conditions. Integrating Eq.~\eqref{eq:programs_beforeint}  over $\tau$, and similarly for the second replisome, directly leads to Eqs.~\eqref{eq:solfi} and \eqref{eq:solfi_params}.

\section{Effect of diffusivity on the shape of the bacterial DNA abundance distribution}\label{ap:curv}

We here study the effect of increasing the diffusivity $D$ on the shape of the DNA abundance distribution $\mathcal{P}(x)$. We first focus on the the meeting point region, and then consider the region far from the expected meeting point. We call the expected meeting point $x=L/2$ the ``replication terminus''. Since this is the last location on the genome to be replicated, $\mathcal{P}$ attains its global minimum there. For $D=0$,  $\mathcal{P}$ exhibits a cusp at the terminus, and thus infinite curvature. For $D>0$, the cusp vanishes, the curvature is finite and decreases with $D$  (Fig.~\ref{fig:bacterial_model}c-d). 

To quantitatively link $D$ with this curvature, we first use that Eq.~\eqref{eq:langevin} is symmetric under the swap of replisomes and mirroring of the genomic coordinate: $f_1(x,
\tau) = f_2(L-x,\tau)$. As a consequence, $f'(L/2)$ vanishes for all time $
\tau$ and therefore $\mathcal{P}'(L/2) = 0$ according to Eq.~\ref{eq:abun1}. The curvature at $x=L/2$ is thus simply $\mathcal{P}''(L/2)$. We focus on the relative curvature expressed by
\begin{equation}\label{eq:A_terminus}
  \frac{\mathcal{P}''(L/2)}{\mathcal{P}(L/2)}
  = \frac{
    \int_0^\infty d\tau \Lambda e^{-\Lambda \tau} f''(L/2,\tau)\
  }{
    \int_0^\infty d\tau \Lambda e^{-\Lambda \tau} f(L/2,\tau)
  }.
\end{equation}
To make progress, we approximate the replication time distribution with a Gaussian:
\begin{equation}
   \psi(x,\tau)= \frac{\partial}{\partial \tau}f(x,\tau) \approx \frac{1}{\sqrt{2\pi\sigma^2}}e^{-\frac{(t-\mu(x))^2}{2\sigma^2(x)}},
\end{equation}
where the mean $\mu(x)$ and the variance $\sigma^2(x)$ have to be determined. It then follows from Eq.~\eqref{eq:abun2} that
\begin{equation}\label{eq:A_gaussian_approx}
    \mathcal{P}(x)= \int_0^\infty d\tau e^{-\Lambda \tau} \psi(x,\tau)
    \approx e^{-\mu(x) \Lambda + \sigma^2(x) \Lambda^2 /2}.
\end{equation}
Here, we have approximated the integral by extending the lower extreme to $-\infty$. We expect the error caused by this approximation to be negligible, since the Gaussian distribution that approximates $\psi(x,\tau)$ must be concentrated on the positive real numbers. By using that $f'(L/2,\tau) \equiv 0$ and thus $\mu'(x) = 0$, and assuming that the dependence of $\sigma$ on $x$ is small enough to be ignored, we obtain
\begin{equation}\label{eq:App_gaussian_approx}
    \frac{\mathcal{P}''(L/2)}{\mathcal{P}(L/2)} \approx -\Lambda \mu''(L/2).
\end{equation}
We now have to evaluate $\mu''(L/2) = \partial_x^2 \int \tau \partial_\tau f(x,\tau) d\tau|_{x=L/2}$. To make this tractable, we now further approximate the individual inverse Gaussian laws of $f_1$, $f_2$ from Eq.~\eqref{eq:solfi} with Gaussian distributions. For these individual laws, the means and variances are given by Eq.~\eqref{eq:solfi_params} which yields
\begin{eqnarray}
    \hat f(x,\tau) = 1 -
    \left[1 - \Phi\left(\frac{\tau - x/v_0}{\sqrt{2Dx/v_0^3}}\right)\right]\times \nonumber \\
    \left[1 - \Phi\left(\frac{\tau - (L-x)/v_0}{\sqrt{2D(L-x)/v_0^3}}\right)\right].
\end{eqnarray}
where we have for simplicity assumed that the replisome speed is constant, i.e. $h(\tau)=1$. Therefore:
\begin{align}
    \mu''(L/2)
    &\approx \left.\frac{\partial^2}{\partial x^2}\int_{-\infty}^\infty \tau \frac{\partial}{\partial \tau} \hat f(x,\tau)d\tau\right|_{x=L/2}
    \nonumber\\
    &= \!\left.\frac{\partial^2}{\partial x^2} \left(
    \int_0^\infty \!\!\!\!\! d\tau \big(1 - \hat f(x,\tau)\big)
    -\!\!\int_{-\infty}^0 \!\!\!\!\!\!\! d\tau \hat f(x,\tau)
     \!\right) \!\right|_{x=L/2}
    \nonumber \\
    &= -\int_{-\infty}^\infty \hat f''(L/2,\tau) d\tau
    \nonumber \\
    &= -\frac{2}{\sqrt{\pi D L v_0}}
\end{align}
and by inserting into Eq.~\eqref{eq:App_gaussian_approx} we obtain Eq.~\eqref{eq:curvature} for the normalized curvature of the DNA abundance at the terminus.

Eq.~\ref{eq:A_gaussian_approx} also permits to approximate the value of $\mathcal{P}(x)$ itself at the terminus. We use that
\begin{align}
    \mu(L/2)
    &\approx \int_{-\infty}^\infty \tau \frac{\partial}{\partial \tau} \hat f(x,\tau)d\tau
    \nonumber \\
    &= \int_0^\infty \big(1 - \hat f(x,\tau)\big)d\tau
     - \int_{-\infty}^0 \hat f(x,\tau)d\tau
    \nonumber \\
    &= \frac{L}{2v_0} - \sqrt{\frac{DL}{\pi v_0^3}}
\end{align}
and 
\begin{align}
    \sigma^2(L/2)
    &\approx \int_{-\infty}^\infty \tau^2 \frac{\partial}{\partial \tau} \hat f(x,\tau)d\tau - \mu^2(L/2)
    \nonumber \\
    &= 2\int_0^\infty \tau\big(1 - \hat f(x,\tau)\big)d\tau
    \nonumber \\
    &- 2\int_{-\infty}^0 \tau \hat f(x,\tau) d\tau
     - \mu^2(L/2)
    \nonumber \\
    &= \frac{\pi -1}{\pi}\frac{D L}{v_0^3}.
\end{align}
Approximating $\mu(L/2+\epsilon) \approx \mu(L/2) + \epsilon^2 \mu''(L/2)/2$ and $\sigma^2(L/2 + \epsilon) \approx \sigma^2(L/2)$ in Eq.~\eqref{eq:A_gaussian_approx} yields
\begin{align}
    \mathcal{P}(L/2+\epsilon) &\approx \exp\Bigg[
    -\frac{\Lambda  L}{2 v_0}
    +\Lambda  \sqrt{\frac{D L}{\pi v_0^3}}
    \nonumber \\
    &\hphantom{\approx \exp\Bigg[}
    +\frac{\pi -1}{\pi}
     \frac{ \Lambda ^2 DL}{2 v_0^3}
    +\epsilon ^2\frac{\Lambda}{\sqrt{\pi D L v_0}}
    \Bigg], \\
\intertext{close to the replication terminus. The curvature at $x=L/2$ is therefore}
    \mathcal{P}''(L/2) &\approx
    \frac{2\Lambda }{\sqrt{\pi D L v_0}}
    \exp\Bigg[
    -\frac{\Lambda  L}{2 v_0}
    +\Lambda  \sqrt{\frac{D L}{\pi v_0^3}}
    \nonumber \\
    &\hphantom{\frac{2\Lambda }{\sqrt{\pi D L v_0}}\approx \exp\Bigg[}
    +\frac{\pi -1}{\pi}
     \frac{ \Lambda ^2 DL}{2 v_0^3}
    \Bigg].
\end{align}

We now study the shape of the DNA abundance far from the terminus region. Within these regions, only one of the two replisomes plays a relevant role. For simplicity, we focus on the region replicated by replisome 1. Also in this case, we assume constant replisome speed, i.e. $h(\tau) = 1$. Since replisome 2 has virtually no chance of reaching the area of interest before replisome 1, we assume $f(x,\tau) \approx f_1(x,\tau)$. The DNA abundance $\mathcal{P}(x)$ can then be explicitly found from the characteristic function of the inverse Gaussian distribution; the expansion $\sqrt {1+x}\approx 1+x/2-x^{2}/8$ yields for small $D$
 \begin{equation}\label{eq:abundance_single_vconst_smallD}
    \mathcal{P}(x) \approx \exp\left[\left(-\frac{\Lambda}{v_0} + D\frac{\Lambda^2}{v_0^3}\right)x\right].
 \end{equation}
The change in the exponential decay rate due to $D$ is therefore of order $D\Lambda^2 / v^3$.

\section{Uncertainty on the meeting point}\label{ap:meetingpoint}

We now consider the random point $Z$ at which the two replisomes meet and its uncertainty $\langle (Z - \langle Z\rangle)^2\rangle$. We call $\tau_1(x)$ and $\tau_2(x)$ the random times at which replisomes 1 and 2 reach location $x$, respectively. Then, $\tau_1(z) < \tau_2(z)$ for all points $z$ to the left of the meeting point $Z$, and $\tau_1(z) > \tau_2(z)$ for all points to the right of $Z$. Therefore
\begin{equation}\label{eq:Pz_1}
    \mathbb{P}(Z < z) =
    \mathbb{P}(\tau_1(z) > \tau_2(z))\ .
\end{equation}
We note that the means and variances in Eq.~\eqref{eq:solfi_params} are the means and variances of $H\big(\tau_i(x)\big)$. Since $H(t)$ increases monotonically, we substitute it into Eq.~\eqref{eq:Pz_1}:
\begin{equation}
    \mathbb{P}(Z < z) =
    \mathbb{P}\big(H(\tau_2(z)) - H(\tau_1(z)) < 0\big).
\end{equation}
By using that $\tau_1(z)$ and $\tau_2(z)$ are independent, and substituting the means and variances from Eq.~\eqref{eq:solfi_params}, we find the mean and variance of $H(\tau_2(z)) - H(\tau_1(z))$ to be $(L - 2z)/v_0$ and $2DL / v_0^3$, respectively. By approximating the distribution of the difference between the two replication times with a Gaussian, we obtain
\begin{equation}
    \mathbb{P}(Z < z)
    \approx
    \Phi\left(\frac{-(L-2z)/v_0}{\sqrt{2DL / v_0^3}}\right)
    = \Phi\left(\frac{z - L/2}{\sqrt{DL / 2 v_0}}\right)
\end{equation}
where $\Phi(x) = 1/2 + \erf\big(x / \sqrt{2}\big)/2$ is the cumulative function of the Gaussian distribution. The uncertainly about the meeting point is therefore given by Eq.~\eqref{eq:uncertainty}.

\section{Bacterial DNA replication with position-dependent speed}\label{ap:vx}

\begin{table}
    \centering
    \begin{tabular}{cccccc}
        $T$ & $\overline{v}$ [$\text{bp}/\text{s}$] & $D$ [$\text{kbp}^2/\text{s}$] &
        $\delta$ & $\omega$ [\text{rad}/\text{Mbp}] & $\phi$ [rad] \\
        \colrule
        17 & 230 $\pm$ 22 & 1.0 $\pm$ 1.2  & 0.18 $\pm$ 0.07 & 6.2 $\pm$ 6.7  & 2.3 $\pm$ 1.6  \\
        22 & 350 $\pm$ 37 & 1.0 $\pm$ 1.6  & 0.23 $\pm$ 0.15 & 2.8 $\pm$ 0.73 & 3.4 $\pm$ 0.72 \\
        27 & 540 $\pm$ 17 & 0.4 $\pm$ 0.5 & 0.18 $\pm$ 0.04  & 4.6 $\pm$ 0.11 & 2.1 $\pm$ 0.11 \\
        32 & 820 $\pm$ 57 & 1.4 $\pm$ 2.3  & 0.12 $\pm$ 0.04 & 5.5 $\pm$ 0.13 & 1.6 $\pm$ 0.16 \\
        37 & 970 $\pm$ 25 & 2.9 $\pm$ 2.3  & 0.18 $\pm$ 0.02 & 4.3 $\pm$ 0.15 & 3.0   $\pm$ 0.14 \\
    \end{tabular}
    \caption{Parameter estimates for position-dependent speed $v(x) = \overline{v}\big(1 + \delta\cos(\omega x + \phi)\big)$ model of Bhat \textit{et al.}\cite{bhat2022speed}.}
    \label{tab:params_vd}
\end{table}

\begin{figure*}
    \hbox to \textwidth{
    \hfill
    \vtop{\hbox to 0in{\ \textbf{(a)}}\hbox{\includegraphics{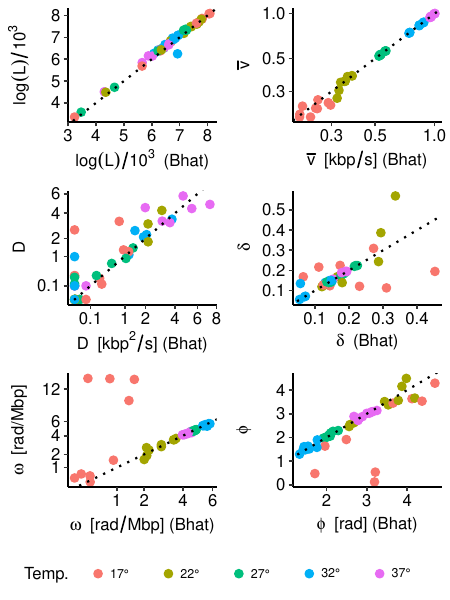}}}%
    \hfill
    \vtop{\hbox to 0in{\ \textbf{(b)}}\hbox{\includegraphics{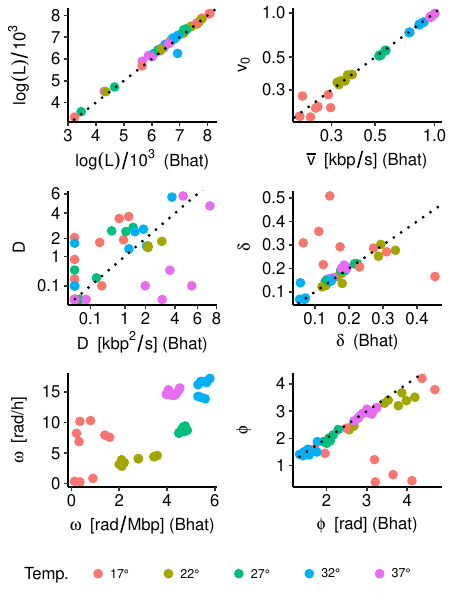}}}%
    \hfill
    }
    \caption{Comparison of our parameter estimates with those of Ref.\cite{bhat2022speed}. The plots shows data points for all 9 combinations of biological replicate (3 per temperature) and stationary reference sample (3 in total, taken at temperature 17$^\circ$C, 27$^\circ$C and 37$^\circ $C). \textbf{(a)}. Position-dependent speed $v(x) = \overline{v}\big(1 + \delta\cos(\omega x + \phi)\big)$. \textbf{(b)}. Time-dependent speed $v(\tau) = v_0\big(1 + \delta\cos(\omega \tau + \phi)\big)$. }
    \label{fig:elifecmp}
\end{figure*}

\begin{figure*}
    \hbox to \textwidth{
    \hfill
    \vtop{\hbox to 0in{\ \textbf{(a)}}\hbox{\includegraphics{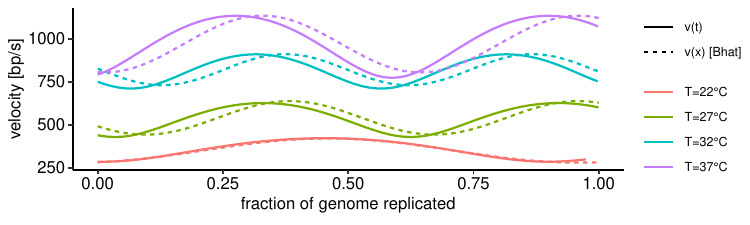}}}%
    \hfill
    \vtop{\hbox to 0in{\ \textbf{(b)}}\hbox{\includegraphics{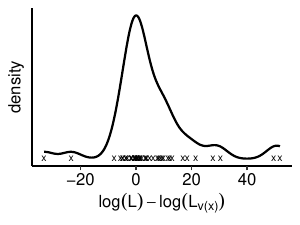}}}%
    \hfill
    }
    \caption{Time-dependent vs. position-dependent speed oscillations. \textbf{(a)}. Time-dependent speed oscillations $v(\tau)$ from figure \ref{fig:elife_data}c vs. the position-dependent speed oscillations from Bhat \textit{et al.}\cite{bhat2022speed}. \textbf{(b)}. Likelihoods $L$ of the best-fitting time-dependent speed model vs. $L_{v(x)}$ of the best-fitting position-dependent model (computed as described in Appendix \ref{ap:parest}).}
    \label{fig:elifecmp_osc}
\end{figure*}

In this Appendix, we solve a model with position-dependent speeds $v_1(x)$ and $v_2(x)$ in the limit of small diffusivity. This will allow us to directly compare time-dependent and position-dependent speed models. We allow the diffusion constants $D_1(x)$, $D_2(x)$ to vary in space as well. The Langevin equations are:
\begin{align}\label{eq:langevin_vx}
  \frac{d}{d\tau} x_1 &= v_1(x) + \sqrt{2D_1(x)}\, \xi_1(\tau) \nonumber\\
  \frac{d}{d\tau} x_2 &= v_2(x) + \sqrt{2D_2(x)}\, \xi_2(\tau).
\end{align}
where $v_1(x) > 0$ and $v_2(x) < 0$. As in the case of Eq.~\eqref{eq:langevin}, the initial conditions are $x_1(0)=0$, $x_2(0)=L$ and we  define replication programs $f_1(x,\tau)$, $f_2(x,\tau)$ via the first-passage times of $x_1(\tau)$ and $x_2(\tau)$, respectively, through position $x$.

We now assume that (i) $v_1(x)$, $v_2(x)$ change only slowly, (ii) $D_1(x)$, $D_2(x)$ are small in comparison, and (iii) $x$ is not too close to the replication origin. Under these assumptions, replication times are approximately additive, meaning that if $\tau_{x_0\to x_1}$ is the (random) time a replisome originating from $x_0$ takes to reach $x_1$, then $\tau_{x_0\to x_1}$ and $\tau_{x_1\to x_2}$ are independent, and $\tau_{x_0\to x_2} = \tau_{x_0\to x_1} + \tau_{x_1\to x_2}$. For large enough $x$, we invoke the central limit theorem for the sum $\tau_x = \tau_{x_0 \to x_1} + \cdots + \tau_{x_k \to x}$ and conclude that $\psi_1$ and $\psi_2$ are approximately Gaussian, implying
\begin{equation}\label{eq:program_gaussian_single_vx}
     f_i(x,\tau) \approx \frac{1}{2} + \frac{1}{2}\erf\left(\frac{\tau - \mu_i(x)}{\sqrt{2}\sigma_i(x)}\right)
 \end{equation}
To find the position-dependent mean $\mu_i(x)$ and variance $\sigma_i^2(x)$, we rely on $v_i(x)$ changing sufficiently slowly relative to $D_i(x)$ so that we can find a subdivision $x_0 \leq \cdots \leq x_k \leq x$ where $v_i(x)$ is approximately constant in each interval. By approximating the mean and variance of $\tau_{x_l \to x_{l+1}}$ with the constant-speed solution, we obtain
\begin{equation}\label{eq:program_gaussian_single_vx_meanvar}
     \mu_i(x) = \int_{x_i(0)}^x \frac{dx'}{v_i(x')},\quad
     \sigma_i^2(x) = \int_{x_i(0)}^x \frac{2D_i(x')}{v_i^3(x')}dx'.
\end{equation}

We now compute $\mu_i$ and $\sigma_i^2$ for the position-dependent speed fluctuations from Ref. \cite{bhat2022speed} of the form 
\begin{equation}\label{eq:harm_v_z}
   v(x) = \overline{v} \left(1 + \delta \cos (\omega x + \phi)\right)
\end{equation}
where $x$ is the distance of the replisome from the origin. We note that because we express $v(x)$ in terms of distance travelled, the speed of the two replisomes when they traverse the same physical location will in general differ. Since integration of $1/v(x)$ and $1/v^3(x)$ is cumbersome, we first compute the following integrals of powers of $g(u) = 1 + \delta\cos u$, which are valid for $z \in (-\pi,\pi)$:
\begin{align}
  \hat G_1(z)
  &= \int_0^z \frac{du}{g(u)}
  = \frac{2 \tanh ^{-1}\left(\frac{(1-\delta) \tan \frac{z}{2}}{\sqrt{1-\delta ^2}}\right)}{\sqrt{1-\delta ^2}}
  \nonumber \\
  \hat G_3(z)
  &= \int_0^z \frac{du}{g^3(u)}
  = \frac{\left(\delta ^2+2\right) \tanh ^{-1}\left(\frac{(1-\delta) \tan \frac{z}{2}}{\sqrt{1-\delta ^2}}\right)}
         {\left(1-\delta ^2\right)^{5/2}}
  \nonumber \\
  &\;\hphantom{= \int_0^z \frac{du}{g^3(u)}}
  + \frac{\delta  \sin (z) \left(\delta ^2-3 \delta  \cos (z)-4\right)}
         {2 \left(1-\delta ^2\right)^2 (\delta  \cos (z)+1)^2}
  \\
\intertext{We also introduce the definite integrals}
  \overline{G}_1
  &= \int_{-\pi}^{\pi} \frac{du}{g(u)} = \frac{2 \pi }{\sqrt{1-\delta ^2}}, \nonumber \\
  \overline{G}_3
  &= \int_{-\pi}^{\pi} \frac{dz}{g^3(z)} = \frac{\pi  \left(\delta ^2+2\right)}{\left(1-\delta ^2\right)^{5/2}}.\\
\intertext{Writing $[[\cdot]]$ for the nearest integer we obtain}
  G_p(z)
  &= \int_0^z \frac{du}{g^p(u)}
  = \hat G_p(z) + \overline{G}_p \bigg[\bigg[\frac{z}{2\pi}\bigg]\bigg]
\end{align}
and in terms of $G_p$ we finally have
\begin{equation}
    V_p(x)
    = \int_0^x \frac{du}{v^p(u)}
    = \frac{v_0}{\omega}\big(G_p(\omega x + \phi) - G_p(\phi)\big).
\end{equation}
By setting $v_1(x) = v(x)$, $v_2(x) = -v(L-x)$ and $D_1(t) = D_2(t) = D$ in Eq.~\eqref{eq:program_gaussian_single_vx_meanvar} and inserting the means and variances into Eq.~\eqref{eq:program_gaussian_single_vx}, we arrive at the replication program
\begin{equation}\label{eq:program_bacteria_harmonic_vd}
  f(x,\tau) = 
  1 - \frac{1}{4}
	 \erfc\left(\frac{t - V_1(x)}{2^{\frac{3}{2}}DV_3(x)}\right)
	 \erfc\left(\frac{t - V_1(L-x)}{2^{\frac{3}{2}}DV_3(L-x)}\right).
\end{equation}

\subsection*{Comparison of parameter estimates}

Most parameter estimates for the position-dependent speed model obtained using the procedure outlined in Appendix \ref{ap:parest} agree well with the values reported in Ref.~\cite{bhat2022speed},  see Fig.~\ref{fig:elifecmp}a. Estimates that deviate also exhibit large uncertainties between replicates, see Table~\ref{tab:params_vd}. These large uncertainties can be explained in a similar way as for the time-dependent model presented in Section~\ref{sec:bacteria}.

Parameter estimates also agree well between time- and position-modulated models, see Fig.\ref{fig:elifecmp}b. The oscillation frequency $\omega$ is not directly comparable between the two models, but when plotted against replication progress both models yield similar oscillations, see Fig.~\ref{fig:elifecmp_osc}a. When the two models are compared across samples, we find that, for a large majority of samples, the time-dependent speed model yields a slightly higher likelihood than the position-dependent speed model, see Fig.~\ref{fig:elifecmp_osc}b. 

\section{Parameter estimation}\label{ap:parest}

\begin{figure*}
\hbox to \textwidth{
\vtop{\hbox to 0in{\ \textbf{(a)}}\hbox{\includegraphics{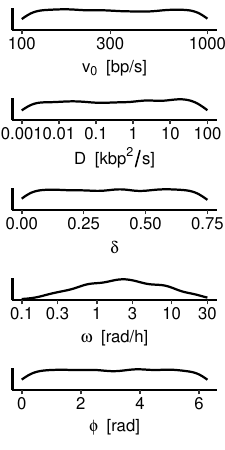}}}%
\hfill
\vtop{\hbox to 0in{\ \textbf{(b)}}\hbox{\includegraphics{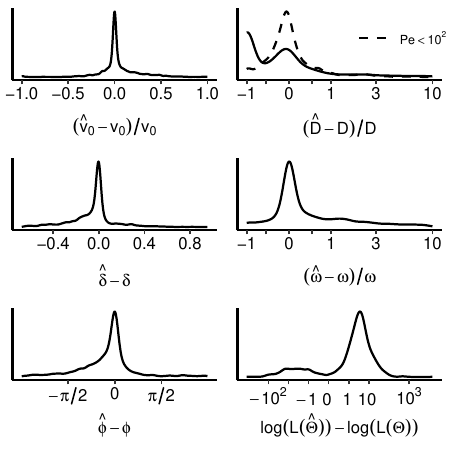}}}%
\hfill
\hbox{\vtop{%
\hbox{\vtop{\hbox to 0in{\ \ \textbf{(c)}}\hbox{\includegraphics{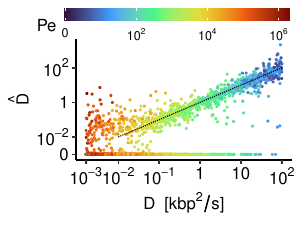}}}}%
\hbox{\vtop{\hbox to 0in{\ \ \textbf{(d)}}\hbox{\includegraphics{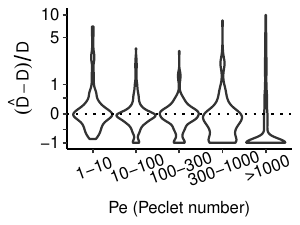}}}}%
}}%
}
\caption{Parameter recovery. \textbf{(a)}. Distributions of true parameters $\theta$. \textbf{(b)}. Estimation errors for simulated abundance profiles with $\Lambda=1\,h^{-1}$ and 40k reads / 10kbp window on average. \textbf{(c)}. True value $D$ vs. recovered value $\hat D$ for different ranges of the Peclet number $\text{Pe} = Lv_0 / 4D$. \textbf{(d)}. Distribution of the relative error of $\hat D$ for different ranges of $\text{Pe}$.}
\label{fig:recovery}
\end{figure*}

We compute the position-wise bias-corrected abundances and their uncertainties as
\begin{equation}\label{eq:log-likelihood-ai-sigmai}
  a_i = \frac{n^\text{exp}_i}{n^\text{stat}_i},\qquad
  \sigma_i = a_i \sqrt{\frac{1}{n^\text{exp}_i}+\frac{1}{n^\text{stat}_i}}
\end{equation}
given the experimentally observed read counts at genomic locations $x_1,\ldots,x_n$ in an exponentially growing ($n^\text{exp}_i$) and stationary ($n^\text{stat}_i$) populations \cite{bhat2022speed}. Since the counts  $n^\text{exp}_i$ and $n^\text{stat}_i$ are typically large, we approximate the error distributions as Gaussian. Defining the parameter vector $\Theta = \big(v_0, D, \delta, \omega, \phi\big)$, we express the log likelihood by
\begin{align}
  \log L(\Theta)
  &= -\frac{N}{2}\log(2\pi) - \sum_{i=1}^N \log(\sigma_i)
  \nonumber \\
  &- \frac{1}{2}\sum_{i=1}^N \frac{\big(a_i - \lambda \mathcal{P}(x_i|\Theta)\big)^2}{\sigma_i^2}.
  \label{eq:log-likelihood}
\end{align}
where $\lambda = \sum_i a_i \big/ \sum_i \mathcal{P}(x_i|\Theta)$ is a scaling factor.

We evaluate $\mathcal{P}(x|\Theta)$  by a numerical integration of Eq.\eqref{eq:abun2}. This procedure requires a limited computational effort compared with Ref.~\cite{bhat2022speed}, that evaluated the likelihood using stochastic simulations.
We fit two different models corresponding to two different replication programs with are both instances of the general bacterial program from Eq.~\eqref{eq:program_bacteria}: (1) the time-dependent speed model in Eq.~\eqref{eq:solfi} with the modulating function $h(\tau)$ given in Eq.\eqref{eq:h_harmonic}. (2) the position-dependent speed model of Ref.~\cite{bhat2022speed} with replication program Eq.~\eqref{eq:program_bacteria_harmonic_vd}, see Appendix \ref{ap:vx} (parameters are $\Theta = \big(\overline{v}, D, \delta, \omega, \phi\big)$ in this case, the parameter $\overline{v}$ takes the place of $v_0$).

The likelihood surfaces of these models are quite rough, in particular with respect to $\omega$ and $\phi$. This causes combined gradient-ascent optimization of all five parameters to remain stuck in local maxima, unless the initial estimate is close to the global optimum.  On the other hand, exhaustive exploration of the parameter space for all 5 parameters at sufficient resolution is computationally expensive. We thus used the following procedure:

\begin{enumerate}
\item We fit the \emph{constant-speed}, \emph{no-diffusion} ($\delta=0$, $D=0$) regime by optimizing $v_0$ while keeping $D=0$ and $\delta=0$. Initial value is $v_0=500 \text{bp/s}$.
\item We fit the \emph{oscillatory} \emph{no-diffusion} ($\delta,\omega,\phi \geq 0$, $D=0$) regime by optimizing $v_0$, $\delta$, $\omega$ and $\phi$ while keeping $D=0$. Initial values lie on a grid formed by the constant-speed, no-diffusion estimate of $v_0$, $\delta \in \{0, 1/2\}$, 25 logarithmically spaced values for $\omega$ from $\pi L /2$ to $32 \pi L$ ($L$ being the genome length), and 5 uniformly spaced values for $\phi$ from $0$ to $2\pi$.
\item We fit the \emph{constant-speed} \emph{diffusive} ($\delta=0$, $D\geq 0$) regime by optimizing $v_0$ and $D$ while keeping $\delta=0$. Initial values lie on a grid formed by the constant-speed, no-diffusion estimate of $v_0$, and 21 logarithmically spaced values for $D$ from $10$ to $10^{10}$.
\item We find the initial parameters for the \emph{oscillatory} \emph{diffusive} ($\delta,\omega,\phi \geq 0$, $D \geq 0$) regime by optimizing $v_0$, $\delta$, $\omega$ and $\phi$ while fixing $D$ to the \emph{constant-speed} \emph{diffusive} estimate. Initial values are the same as in (2), except that for $v_0$ we now use the \emph{constant-speed} \emph{diffusive} estimate.
\item We fit the \emph{oscillatory} \emph{diffusive} regime by optimizing $v_0$, $D$, $\delta$, $\omega$ and $\phi$, using the estimates found in (1)-(4) as initial values.
\item We select the parameter regime with the lowest AIC score.
\end{enumerate}

\subsection*{Identifiability}

To study the identifiability of parameters $v_0$, $D$, $\delta$, $\omega$, $\phi$ of the bacterial replication program Eq. \ref{eq:solfi} with $h(t) =  v_0\big(1+\cos(\omega t + \phi)\big)$ we generated artificial DNA abundance distributions and estimated the parameters values using the procedure outlined in Appendix~\ref{ap:parest}. We randomly selected 2000 sets of parameters (Fig.~\ref{fig:recovery}a), and computed their predicted abundance profiles by simulating $10^6$ genomes using a fixed growth rate of  $\Lambda = 1\text{h}^{-1}$). We then generated Poisson-distributed read counts $n^\text{exp}_i$ (and similarly for $n^\text{stat}_i$, using a flat abundance profile) from these abundance profiles, and ran the estimation procedure described in appendix \ref{ap:parest}. All parameters except $D$ were recovered well (Fig.~\ref{fig:recovery}b). 

Diffusivities $D$ below a certain value are unlikely to pass the Akaike information criterion and are therefore estimated to be zero. This occurs frequently for parameter sets corresponding to a Peclet number $\text{Pe} = Lv_0 / 4D$ larger than 300 (Fig.~\ref{fig:recovery}c-d). 

\section{Eukaryotic DNA replication}\label{ap:eukaryotes}

In this Appendix, we derive Eq.~\eqref{eq:abundance_eukaryotic} for the DNA abundance in the case of origins at well-defined locations and with time-homogeneous firing rates, see Eq.~\eqref{eq:yeast_Ixt}. The general eukaryotic replication program is given by 
\begin{equation}\label{eq:fxt_euk}
f(x,\tau)=1-e^{-\int\int dx'd\tau' I(x',\tau') \theta\left(\tau-\tau'-\frac{|x-x'|}{v}\right)}\, ,
\end{equation}
see \cite{bechhoefer2012replication,baker_inferring_2014}. Equation~\eqref{eq:fxt_euk} is equivalent to Eq.~\eqref{eq:program_cone}, but with an explicit expression for the past light cone. We note that
\begin{gather}
    \frac{d}{d\tau} \int\int dx'd\tau' I(x',\tau')\theta\left(\tau-\tau'-\frac{|x-x'|}{v}\right)
    \nonumber \\
    =\int dx' I\left(x',\tau-\frac{|x-x'|}{v}\right) .
\end{gather}
We use Eq.~\eqref{eq:abun1} to obtain
\begin{gather}
\mathcal{P}(x)=\int_0^\infty dt\,
\left[ \int dx'' I\left(x'',\tau-\frac{|x-x''|}{v}\right) \right]\times
\nonumber \\
\left[e^{-\Lambda \tau-\int\int dx'dt' I(x',\tau') \theta\left(\tau-\tau'-\frac{|x-x'|}{v}\right)}\right].
\end{gather}
Setting $I(x,\tau)=\sum_{j=1}^K I_j(\tau)\delta(x-x_j)$, i.e. assuming discrete origins $x_1,\ldots,x_K$, yields
\begin{gather}
    \mathcal{P}(x)=\int_0^\infty d\tau\, \left[\sum_j I_j\left(\tau-\frac{|x-x_j|}{v}\right)\right]\times
    \nonumber \\
	\left[e^{-\Lambda \tau-\sum_j\int_0^{\tau-|x-x_j	|/v} d\tau' I_j(\tau')}\right],
\end{gather}
and in the time homogeneous case $I_j(\tau)=I_j^*\theta(\tau)$ we find
\begin{eqnarray}
    \mathcal{P}(x)=1{-}\Lambda
    \!\!\int_0^\infty \!\!\!\!\!\!\! d\tau\, e^{-\Lambda \tau-\sum_j\!I_j^* \big(\tau-\frac{|x-x_j|}{v}\big)\theta\big(\tau-\frac{|x-x_j|}{v}\big)}.
\end{eqnarray}
The integral can be then done by breaking it into different time intervals in which only a certain set of the theta functions are non-zero. This yields
\begin{eqnarray}
&& \int_0^\infty d\tau\, e^{-\Lambda \tau-\sum_j I_j^* (\tau-|x-x_j|/v)\theta(\tau-|x-x_j|/v) }\nonumber\\
 &&=\frac{1-e^{-\Lambda \tau_1}}{\Lambda}
 \nonumber \\
 &&+\sum^{K-1}_{k=1}\frac{e^{-\Lambda \tau_k-\sum^{k}_{j=1}I^*_j(\tau_k-\tau_j)}-e^{-\Lambda \tau_{k+1}-\sum^{k}_{j=1}I^*_j(\tau_{k+1}-\tau_j)}}{\Lambda+\sum_{j=1}^k I^*_j}
 \nonumber \\
 &&+\frac{e^{-\Lambda \tau_K-\sum^{K}_{j=1}I^*_j(\tau_K-\tau_j)}}{\Lambda+\sum_{j=1}^K I^*_j}
\end{eqnarray}
where $\tau_k=\frac{|x-x_k|}{v}$ are the times it takes a replisome formed at the $k$-th origin to reach position $x$, and are ordered for g    iven $x$ in such a way that $0 \leq \tau_1 \leq \cdots \leq \tau_K$. Therefore, we have
\begin{eqnarray}
    && \mathcal{P}(x)
    =e^{-\Lambda \tau_1}
    \nonumber \\
    &&-\Lambda\sum^{K-1}_{k=1}\frac{e^{-\Lambda \tau_k-\sum^{k}_{j=1}I^*_j(\tau_k-\tau_j)}-e^{-\Lambda t_{k+1}-\sum^{k}_{j=1}I^*_j(\tau_{k+1}-\tau_j)}}{\Lambda+\sum_{j=1}^k I^*_j}
    \nonumber \\
    &&-\Lambda\frac{e^{-\Lambda \tau_K-\sum^{K}_{j=1}I^*_j(\tau_K-\tau_j)}}{\Lambda+\sum_{j=1}^K I^*_j}.
\label{eq:abundance_eukaryotic_long}
\end{eqnarray}
Equation~\eqref{eq:abundance_eukaryotic} is obtained from Eq.~\eqref{eq:abundance_eukaryotic_long} by substituting $\mathcal{T}_k$ and $W_k$ as defined in Eqs.~\eqref{eq:eukaryotes_Ti} and \eqref{eq:eukaryotes_Wi} and grouping terms with the same exponent.

\section{Eukaryotic origin inference by simulated annealing}\label{ap:simann}

We fit the solution of our eukaryotic model, Eq.~\eqref{eq:abundance_eukaryotic}, to experimentally measured DNA abundances in an exponentially growing asynchronous population of budding yest (\textit{S. cerevisiae} W303)\cite{muller_2013}. Also in this case, we correct for sequencing bias using the DNA abundance measured in a stationary population. Our input data $a_1,\ldots,a_N$ are thus the ratios of number of reads in the exponential vs. stationary population found within each 1kb window along the W303 genome (GenBank accession \textit{CM007964.1}). To estimate number and positions of origins and their fork firing rates, we use simulated annealing \cite{bertsimas_1993}. 

Our free parameters are the number of origins $K$; their positions $x_j$ and intensities $I_j^*/v_0$, with  $j=1\dots K$; and the ratio of growth rate to speed $\Lambda/v_0$. We collectively refer to the vector of $2K+1$ free parameters (excluding $K$ itself) as $\Theta$. To prevent over fitting, we use the Akaike information criterion (AIC) as the cost function, defined as 
\begin{equation}\label{eq:AIC}
    \rm{AIC}(\Theta)=2(2K+1) - 2\log L(\Theta).
\end{equation}
The first term on the right hand side of Eq.~\eqref{eq:AIC}
penalizes a large number of origins, while the second term penalizes small log-likelihoods. The quantity $\log L(\Theta)$ is computed as in Eq.~\eqref{eq:log-likelihood}. The noise in the data appears larger than expected by Poissonian sampling. We therefore use the empirical noise estimate $\sigma_i = \sigma = \sqrt{\frac{1}{N-1} \sum_i (a_{i+1}-a_i)^2 / 2} \approx 3.3\cdot 10^{-9}$ instead of the Poissonian error estimates from Eq.~\eqref{eq:log-likelihood-ai-sigmai}. Our estimate is based on the assumption that the systematic contribution to the quantity $\sigma$ is negligible compared to the noise. We confirmed post-hoc this assumption by replacing the $a_i$ with the fitted model predictions. The resulting value $2\cdot 10^{-10}$ is an order of magnitude smaller; $\sigma$ therefore indeed mainly represents measurement errors rather than systematic effects. 

Our simulated annealing algorithm comprises the following steps:
\begin{enumerate}
    \item We set the Monte-Carlo temperature to $T_m=200$ and generate an initial parameter vector $\Theta$. We draw $\Lambda/v_0$ uniformly from [$7.7 \times 10^{-8}$, $2.3 \times 10^{-5}$] ${\rm bp^{-1}}$ and place two origins randomly on each chromosomes with intensities $I_j^\star / v_0$ drawn uniformly from [$1.1 \times 10^{-7}$, $ 1.6 \times 10^{-3}$] $~{\rm bp^{-1}}$.
    \item We generate a parameter proposal $\Theta'$ by executing one of the following moves with the indicated probability:\\
    \begin{itemize}
        \item \emph{Relocating an origin (prob. 24\%).} We choose one among the $K$ origins randomly and relocate it to a uniformly drawn position on the same chromosome. The origin firing rate remains unchanged.
        \item \emph{Altering one origin intensity (prob. 24\%).} We randomly choose one of the origins and replace its intensity rate $I_j^\star / v_0$ with a value drawn uniformly from [$1.1 \times 10^{-7}$, $ 1.6 \times 10^{-3}$] $~{\rm bp^{-1}}$. 
        \item \emph{Removing an origin (prob. 24\%).} We choose one among the $K$ origins and remove it unless it is the only origin on the chromosome.
        \item \emph{Adding an origin (prob. 24\%).} We add an origin at a uniformly drawn location on the genome with an intensity  uniformly drawn from [$1.1 \times 10^{-7}$, $ 1.6 \times 10^{-3}$] $~{\rm bp^{-1}}$.
        \item \emph{Altering the growth rate (prob. 4\%).} We draw a new value of $\Lambda/v$ uniformly from [$7.7 \times 10^{-8}$, $2.3 \times 10^{-5}$] ${\rm bp^{-1}}$.  
    \end{itemize}
    The ranges of free parameters are chosen to cover replisome speeds ($v$) from $10 {\rm bps^{-1}}$ to $100 {\rm bps^{-1}}$, doubling times $\ln 2/ \Lambda$ between 50 minutes to 150 minutes, and fork firing delays $1/I_j^*$ between 1 minute and 150 minutes.
    \item We accept $\Theta'$ with Metropolis probability
    \begin{equation}
q={\rm min}\left\{1,~\exp\left(\frac{\rm{AIC}(\Theta)-\rm{AIC}(\Theta')}{T_m}\right) \right\}\, .
\end{equation} 
\item We reduce the Monte-Carlo temperature by a factor  $1-1.336\cdot 10^{-5}$ and reiterate from step (2). The total number of iterations is $2\cdot 10^6$. With this choice, the final temperature is equal to $T_m=5\cdot 10^{-10}$.
\end{enumerate}

After half the number of iterations ($10^6$), the number of origins is already close to the final value (222 vs. the final number of 234), and repeating the procedure 10 times yield similar final numbers of origins (225 $\pm$ 6). We take this as evidence that the algorithm has  converged after $2\cdot 10^6$ iterations.

\bibliography{replication}

\end{document}